\documentclass{jfm}

\usepackage{hyperref}
\hypersetup{
    colorlinks,
    linkcolor={red},
    citecolor={blue},
    urlcolor={blue}
}
\usepackage{graphicx}
\usepackage{epstopdf, epsfig}
\usepackage{amsmath}
\usepackage{fancyhdr}
\usepackage{graphicx}
\usepackage{caption}
\usepackage{subcaption}
\usepackage{mathrsfs}
\usepackage{float}
\restylefloat{table}
\usepackage{wasysym}
\usepackage{soul}
\usepackage{pifont}

\newcommand{\cmark}{\ding{51}}%
\newcommand{\xmark}{\ding{55}}%

\shorttitle{Turbulence statistics
in a negatively buoyant plume}
\shortauthor{Bordoloi and others}

\title{Turbulence statistics
in a negatively buoyant multiphase plume}

\author{Ankur D. Bordoloi\aff{1}
  \corresp{\email{ankur.bordoloi@berkeley.edu}},
   Chris C.K. Lai\aff{2}
  Laura Clark\aff{3}
  Gerardo Veliz\aff{4}
  \and Evan Variano\aff{1}}

\affiliation{\aff{1}Department of Civil and Environmental Engineering, University of California Berkeley, Berkeley, CA 94720, USA
\aff{2}Physics Division, Los Alamos National Laboratory, Los Alamos, NM, 87545
\aff{3}Department of Civil and Environmental Engineering, Stanford University, Stanford, CA 94305, USA
\aff{4} Department of Civil and Environmental Engineering, Cornell University, Ithaca, NY 14853

}

\begin{document}

\maketitle

\begin{abstract}
We investigate the turbulence statistics in a {multiphase plume made of heavy particles (particle  Reynolds number at terminal velocity is 450)}. Using refractive-index-matched stereoscopic particle image velocimetry, we measure the locations of particles {whose buoyancy drives the formation of a multiphase plume,}  {together with the local velocity of the induced flow in the ambient salt-water}. {Measurements in the plume centerplane exhibit self-similarity in mean flow characteristics consistent with classic integral plume theories.} The turbulence characteristics resemble those measured in a bubble plume, {including strong anisotropy in the normal Reynolds stresses. However, we observe structural differences between the two multiphase plumes. First, the skewness of the probability density function (PDF) of the axial velocity fluctuations is not that which would be predicted by simply reversing the direction of a bubble plume. Second, in contrast to a bubble plume, the particle plume has a non-negligible fluid-shear production term in the turbulent kinetic energy (TKE) budget. Third, the radial decay of all measured terms in the TKE budget is slower than those in a bubble plume.}  Despite these dissimilarities, a bigger picture emerges that applies to both flows. The TKE production by particles (or bubbles) roughly balances the viscous dissipation, except near the plume centerline. The one-dimensional power-spectra of the velocity fluctuations show a -3 power-law that puts both the particle and bubble plume in a category different from single-phase shear-flow turbulence.

\end{abstract}
\section{Introduction}

Plumes containing bubbles, particles and droplets are present in both environmental and industrial applications.  A few examples of environmental interest are settling sediment, volcanic eruption columns, CO$_2$ ocean sequestration plumes, and  rising oil droplets and gas bubbles from oil well blowouts  \citep{Freeth1987, Baines2005, Baines2008, Socolofsky2008, Woods2010, Socolofsky2011,Huppert2014, Wang2016}. Suspension-flow plumes differ from traditional single-fluid plumes in that the energy due to buoyant forcing is transmitted indirectly from the suspended phase to the continuous phase. The relative motion between the two phases introduces additional length and time scales; these must be included in the model formulation employed to predict plume behavior, for example, when extending classic single-phase integral plume models  \citep{Morton1956} to multiphase plumes \citep{Milgram1983, Sun1986}.  Identifying such scales is nontrivial due to the complexity of particle-particle and particle-fluid interactions. 

For the special case of air bubbles in water, empirical data collection has allowed accurate closure of predictive schemes \citep{Lance1991, Risso2018, Risso2002,  Mercado2010, Almeras2017}.   However, the bubble-in-water plume can be quite different from other suspension-flow plumes of interest, because bubbles are far less dense than the ambient fluid (specific gravity of order 10$^{-3}$) and have negligible inertia.  Other plumes of interest are droplets in water (specific gravity of order 1), solid particles in water (specific gravity of order 2 to 10), or liquid droplets in air (specific gravity of order 10$^3$). Each of these different suspensions can behave quite differently than the others in terms of the interphase interactions. Empirical data are relatively scarce for these other suspension-flow plumes.

Recent progress in suspension flows (especially turbulent flows) offers the hope that predictive techniques will eventually describe overall plume behavior from a direct description of the internal dynamics of particle-fluid coupling.  To help support this strategy and address some of the open questions that remain about turbulent suspension flows \citep{Balachandar2010, Guazzelli2012}, we take an observational approach herein, measuring the particle and fluid behaviors in a  particle-laden plume. Because particles are dynamically quite different from air bubbles in water, we are curious to see how the two-phase kinematics and the interstitial fluid turbulence behave.  

Both bubbles and particles have been shown to modulate the turbulent properties of multiphase flows in a manner sensitive to volume fraction ($\phi$), particle Reynolds number ($Re_d$), and particle Stokes number ($St$).  One starting point for understanding such turbulence modulation is the agitation of a quiescent fluid by bubbles or particles distributed homogeneously in space.  This is often referred to as \emph{pseudo-turbulence} (\emph{e.g.}  \citet{Lance1991, Cartellier2001}) as its characteristic scales are set by the bubble or particle wakes.  Pseudo-turbulence is fundamentally anisotropic, with stronger velocity fluctuations aligned in the direction of suspended-phase motion ($x_1$, typically vertical). Numerical simulations of pseudo-turbulence by \citet{Riboux2013} show that the velocity fluctuation energy spectrum has a slope of $k^{-3}$ for wavelengths smaller than the integral length scale; they also find that the integral length scale is the ratio of the bubble or particle diameter and the drag coefficient, {C$_d$}.

A recent topical review article by \citet{Risso2018} provides a comprehensive summary of pseudo-turbulence due to a homogeneous swarm of bubbles rising in a quiescent medium. First, the turbulence intensity is anisotropic and is dominated by the vertical component. The probability density function (PDF) of the vertical component is positively skewed due to the wake immediately behind each rising bubble, whereas the other two components are symmetrically distributed with exponential tails.  Because the turbulent kinetic energy (TKE) production in suspension flows occurs at two length scales (first at the scale of bubble wakes and second at the scale of overall bubble population), the combination of the two results in a $k^{-3}$ spectral subrange.

To our knowledge, there are no existing studies examining turbulence in negatively buoyant particle plumes released into an unstratified, initially quiescent fluid.  However, there are many studies of turbulence in bubble plumes with notable contributions coming from \citet{Soga2004, Wain2005, Garcia2006, Seol2009, Simiano2009, Lai2019}. These investigations reveal different velocity characteristics at different axial distance ($x_1$) from the origin. \citet{Simiano2009} examined the near-field characteristics {($x_1/D<2$, where $D$ is the dynamic length scale to be defined in section \ref{sc:mean})} and showed that the meandering nature of bubbles influences the effective spreading, volume fraction, and mean velocity profiles.  \citet{Seol2009} and \citet{Lai2019} focused on the far field ($2<x_1/D<11$) turbulence characteristics, such as Reynolds-stress, turbulent transport and TKE budget; their results will be used to compare with our particle plume data. 

Characterizing the turbulence statistics in a multiphase plume is challenging primarily due to the difficulty in simultaneously measuring velocity in both phases. Traditional intrusive techniques such as hotwire anemometry suffer from the possibility of being damaged by the solid particles. Optical measurements from techniques such as particle image velocimetry (PIV) and laser doppler anemometry are usually distorted due to the difference in optical properties of the two media. We overcome the distortion herein by carefully choosing two media with matched refractive indices.  

This paper is organized as follows: section \ref{sc:exp} provides a detailed description of the plume-generating facility and method for characterizing the two-phase flow. We report and discuss our experimental observations in section \ref{sc:results}. {Finally, section \ref{sc:conclusions} provides the key conclusions from this work.}

 
\section{Experimental setup and methodology}
\label{sc:exp}
\subsection{Plume facility}
The plume experiments were conducted in a rectangular tank (80 cm deep $\times$ 80 cm wide $\times$ 365 cm long) as described in \citet{Bordoloi2017}. The tank was filled with tap water filtered to 5 microns and maintained using an UV purifier. 91.1 kg of commercial sodium chloride (Cargill Top-Flo) was then mixed to yield a salt concentration of 0.04 g/mL. The resulting solution has density $\rho_s$ = 1.04 $\mathrm{g/cm^3}$ and kinematic viscosity $\mu_s$ = 1.059$\times10^{-3}$ Pa.s (see Table \ref{tab:1}).

\begin{figure}
\centering
	 \begin{subfigure}[b]{0.8\textwidth}
        \includegraphics[width=\textwidth]{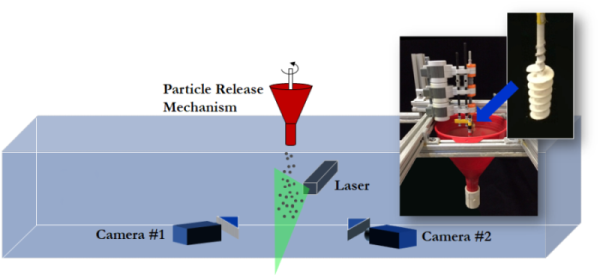} 
         \caption{}
        \label{fig:setup}
        \end{subfigure}

    \begin{subfigure}[b]{0.8\textwidth}
        \includegraphics[width=\textwidth]{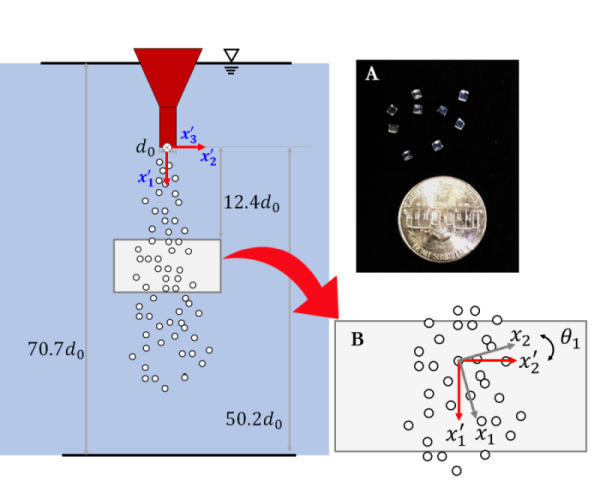} 
         \caption{}
        \label{fig:schematic}
        \end{subfigure}
 \caption{a) Schematic of experimental facility and plume release mechanism (in inset), b) regions of interest with specific dimensions, a picture of teflon particles (inset A) and an illustration of two-dimensional transformation from world coordinates into the plume coordinates (inset B).}
\end{figure}

\begin{table}
  \begin{center}
\def~{\hphantom{0}}
  \begin{tabular}{lcc}
     \textbf{Parameter}             & \textbf{Particle }              & \textbf{Fluid} \\
                                 & \textit{(Teflon PFA)}              & \textit{(0.4\% salt water solution)} \\[3pt]
	Length (mm)      & 2.0                   & --\\
  Bulk volume (mL)  & 110                  & 2.28 $\times10^{6}$\\
  Density ($\mathrm{g/cm^3}$)     & 2.1                    & 1.04\\
  Dynamic viscosity (kgm$^{-1}$s$^{-1}$)& --                 & 1.059$\times10^{-3}$ \\
Refractive index       & 1.34                    & 1.338
  \end{tabular}
  \caption{Physical properties of particles and surrounding salt-water solution.}
  \label{tab:1}
  \end{center}
\end{table}

A schematic of the experimental setup is shown in figures \ref{fig:setup}. The negatively buoyant plume was created by releasing 110 mL of cylindrical PFA pellets (Neoflon AP-202) from a height of 56.5 cm via a screw-feed particle release mechanism (see figure  \ref{fig:schematic}).  The pellets are right circular cylinders with length=diameter=2 mm (see figure \ref{fig:schematic}-A). The physical properties of the solid particles and the surrounding salt solution are summarized in Table \ref{tab:1}. Because of the hydrophobic nature of PFA, the particles tend to trap and hold air bubbles when added to water.  To prevent these air bubbles from entering the experiment, particles were presoaked in water in a separate container and rapidly stirred to dislodge all bubbles. Once the particles were free of air bubbles, they were placed in a funnel for eventual release into the quiescent salt-water mixture.  The funnel was kept partially submerged 23 cm below the free surface through a nozzle with internal diameter $d_0$ = 11.25 mm so that the particles did not contact air (see figure \ref{fig:schematic}). Particle release was governed by a motor-driven helical screw. Prior to release, the particles were held inside the funnel by the blades of the screw. Upon release, the motor rotated the screw at a constant rate of 0.5 RPM operated via Lego Mindstorms software. The particle flux, $Q_0$ = 7.5 cm$^3$/s was measured by video-recording the release of a known quantity of particles and measuring the time difference between the exit of the first and last particle from the funnel nozzle. Before each experiment, the tank fluid was seeded with optical tracers, specifically 13 $\mu$m silver-coated hollow glass spheres (SH400S20, Conduct-O-Fil, Potters Industries). Twenty plume releases provided a sufficient number of samples for the analysis described in Section \ref{sc:results}.


\subsection{Refractive index matching}
\begin{figure}
\centering
        \begin{subfigure}[b]{0.43\textwidth}
        \includegraphics[width=\textwidth]{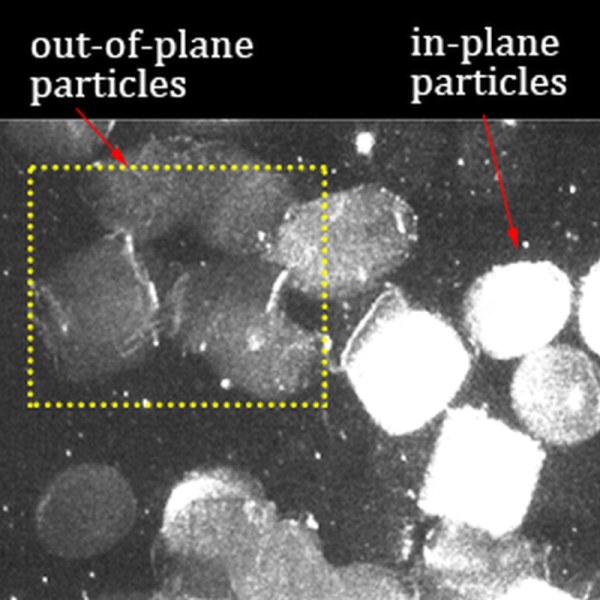} 
          \caption{}
        \label{fig:RIMa}
        \end{subfigure}
        \begin{subfigure}[b]{0.43\textwidth}
        \includegraphics[width=\textwidth]{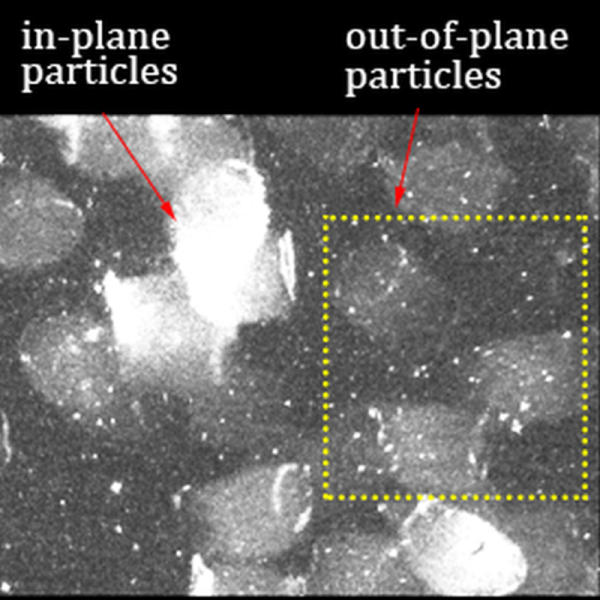} 
          \caption{}
        \label{fig:RIMb}
        \end{subfigure}
	 \caption{Teflon particles (large) and silver coated hollow glass spheres (small) illuminated by a 532 nm laser sheet in a) pure water b) 0.4\% salt water mixture.  }\label{fig:RIM}
\end{figure}

To measure the fluid velocity inside and around the plume, we matched the refractive index (\emph{n}) of the particle and fluid phases. A target refractive index of PFA (\emph{n} $\approx$ 1.34) was achieved from an aqueous solution of commercial sodium chloride with a concentration of 0.04 g/mL. The refractive index of the solution was measured to be 1.338 using an ATAGO refractometer and found to match with the salinity vs. \emph{n} prediction given in \citet{Tan2015}. Although the refractive indices of the two phases were matched, because of the scattering properties of PFA the particles appeared bright under the laser illumination. To limit the intensity of particles below the saturation threshold of the camera's sensor we used a circular polarizer on each camera lens.  Figure \ref{fig:RIM} shows example images of Teflon particles in different-salinity water samples illuminated by a laser sheet (wavelength=532 nm). The high-intensity sections in the images are the particles intersected by the laser sheet, and the low-intensity sections are the out-of-plane particles. Fluid tracers are much more visible when particles and fluid have matched refractive indices (figure \ref{fig:RIMb} compared to figure \ref{fig:RIMa}), especially behind or in front of out-of-plane particles as shown in the highlighted regions.


\subsection{Multiphase velocimetry}
We performed stereoscopic particle image velocimetry (SPIV) to find the two-dimensional three-component (2D3C) velocity field of the fluid phase of the plume. The origin of our coordinate system is located at the tank center. The plume is axisymmetric about the vertical axis, but we use Cartesian coordinates ($x_1$, $x_2$, $x_3$) for our measurements as described in figure \ref{fig:schematic}.  Velocity measurement of the particle phase is currently underway and beyond the scope of this paper.  

The $x_3$ = 0 plane was illuminated with a laser sheet (1 mm thick at beam waist; Quantel/Big Sky Lasers, 532-nm double-pulse Nd-YAG).  Two charge-coupled device (CCD) cameras (Imager PRO-X, 1600 pixels $\times$ 1200 pixels) were synchronized with the laser pulses. They were placed $\pm 55^\circ$ to the laser sheet (c.f. 90$^\circ$ for standard 2D2C PIV). To minimize distortion due to this angle, water-filled acrylic prisms were placed between the camera lenses and the tank walls. The cameras were mounted with Nikkor 105 mm lenses, circular polarizers, and  Scheimpflug adapters.  The interframe delay ($\Delta$t) was optimized to 0.5 ms. The PIV images were collected at a frequency of 14.0 Hz.  {Both fluid and particle motions were decorrelated at timescales $>$ 1/14 s. This setup gave approximately 50 independent} samples during the steady-state phase of each experimental run.

\subsubsection{Fluid-phase processing}
\begin{figure}
\centering
 \includegraphics[width=0.9\textwidth]{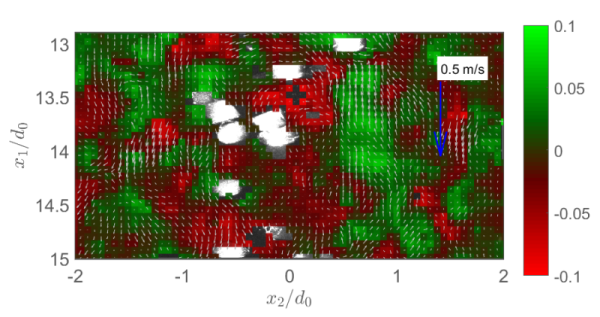}
\caption{Sample mean-subtracted instantaneous two-dimensional-three-component (2D3C) velocity field in the fluid phase superposed with the particle image in the laboratory coordinate system. The vectors show the in-plane velocity ($u_1$, $u_2$, m/s), and the shades of red and green colors show the out-of-plane velocity component ($u_3$, m/s).  The reference vector on the upper right corner corresponds to 0.5 m/s.}
\label{fig:sample}
\end{figure}

We computed the fluid-phase velocity field using DaVis 8.2 software (LaVision GmbH). Before stereo-PIV processing, we performed stereo self-calibration to reduce disparity in the alignment between the laser sheet and the measurement plane to below 0.1 pixel. Tracers and particles in an image were separated by intensity thresholding; erosion and dilation were applied to isolate individual particles. Fluid velocity fields were obtained by masking particles and correlating tracer locations through multipass stereoscopic cross-correlation. The cross-correlation was applied with an initial interrogation window of 64 $\times$ 64 pixels and final window of 48 $\times$ 48 pixels with 50\% overlap, yielding a spacing between vectors of 0.7 mm. {Interrogation windows were weighted according to the symmetric two-dimensional Gaussian function. Vectors were discarded based on universal outlier detection \citep{Westerweel2005} and left as data gaps.} Figure \ref{fig:sample} shows a representative field of fluid velocity fluctuations superposed with the corresponding particle image (transformed into the lab coordinate system). The grid cells are color-coded with the out-of-plane velocity component while the vectors show the in-plane velocity field.

\subsubsection{Particle-phase processing}
\label{sc:particlephase}
\begin{figure}
\centering
\begin{subfigure}[b]{0.48\textwidth}
        \includegraphics[width=\textwidth]{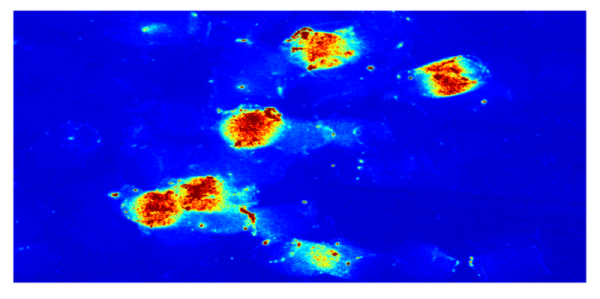} 
          \caption{}
        \label{fig:imp0}
        \end{subfigure}
    \begin{subfigure}[b]{0.48\textwidth}
        \includegraphics[width=\textwidth]{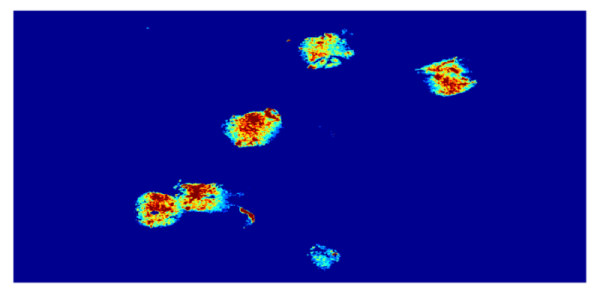} 
          \caption{}
        \label{fig:imp1}
        \end{subfigure}
		\begin{subfigure}[b]{0.48\textwidth}
        \includegraphics[width=\textwidth]{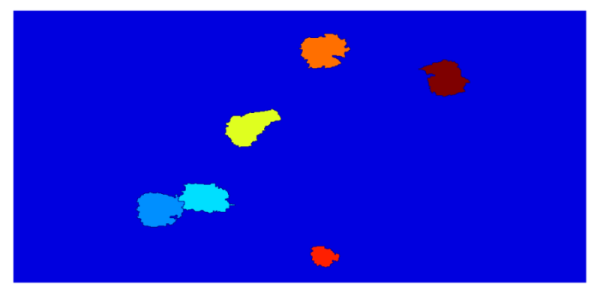} 
          \caption{}
        \label{fig:imp2}
        \end{subfigure}
	  \begin{subfigure}[b]{0.48\textwidth}
        \includegraphics[width=\textwidth]{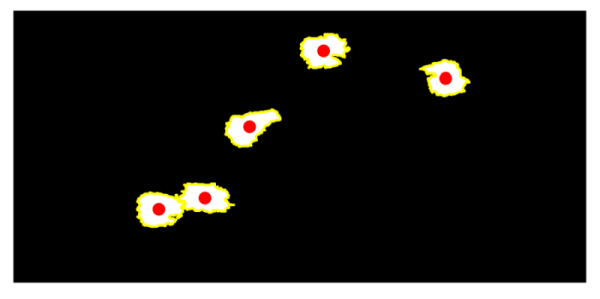} 
          \caption{}
        \label{fig:imp3}
        \end{subfigure}

    \caption{Image processing steps leading from raw PIV image to centroid identification: sample image showing a) particles from Camera 1 and Camera 2 blended into one image, b) convolution of Camera 1 image with Camera 2 image, c) image segmentation based on watershed transform, and d) final binary image with particle boundaries and located centroids in the  laboratory coordinate system.  }\label{fig:improc}
\end{figure}

To estimate the particle number density across the plume, we conducted a series of additional image-processing steps (see figure \ref{fig:improc}). First, the raw images were transformed from the camera coordinate system to the lab coordinate system using the stereoscopic mapping function. Small-size background noise features (including tracers) were removed using a simple median filter. 

In stereo-mapping, after transforming into the lab coordinate system, the section of the particle intersected by the laser sheet should overlap in both cameras. Figure \ref{fig:imp0} shows a sample instant with the two camera images overlayed. The high intensity regions in the image represent the intersection between the two camera images, whereas the low intensity background is from the out-of-plane particles captured in only one of the two cameras. We remove the background noise by setting an intensity threshold and convolving the two images. The result is shown in figure \ref{fig:imp1}. 

After inspecting our dataset, we find many instances where particles are nearly touching as exemplified in the lower left pair in the image. The  nonuniform intensity gradients in the bordering regions among different particle pairs yielded different intensity gradients. A traditional intensity-gradient-based image segmentation therefore could not differentiate between two particles, and so adopted a segmentation technique based on the {watershed transform}. The watershed transform treats an image as a topographic map with brighter intensity pixels as heights, and finds the separating line that runs along the ridges \citep{Gonzalez2007}. The result, with individual particles identified with separate colors, is shown in figure \ref{fig:imp2}. We applied an area threshold to remove partially illuminated particles and clusters that our method could not isolate. The centroids of particles identified from the final binary image are shown in figure \ref{fig:imp3}.

\subsection{Coordinate transformation}

Typically in jets and plumes the cross-stream velocity components ($U_{2}$ and $U_{3}$) are smaller  than the axial velocity component ($U_{1}$) near the centerline. Thus, any small misalignment between the plume axis and the measurement axis could lead to a systematic bias in the measured $U_{2}$ and $U_{3}$. The scenario is schematically shown in inset B in figure \ref{fig:schematic} in a simplified $x_1-x_2$ plane in which the plume axis ($x_1$) makes an angle $\theta_1$ with the measurement axis $x_1^{\prime}$. For our data, if $\theta_1$ is defined the same way and if $\theta_2$ is the angle between the plume axis and the measurement axis on the $x_1-x_3$ plane, using axisymmetry and assuming that $\left<U_{2}\right>, \left<U_{3}\right> = 0$ at the plume centerline, the measured velocities can be transformed into the plume co-ordinate system via correction angles,

\begin{equation}
 \theta_1 = atan \left(\frac{\left<U_{2}^{\prime}\right>_0}{\left<U_{1}^{\prime}\right>_0}\right); 
\theta_2 = atan \left(\frac{\left<U_{3}^{\prime}\right>_0}{\left<U_{1}^{\prime}\right>_0}\right).
\label{eq:trans_angle}
\end{equation} 

\begin{figure}
\centering
    \begin{subfigure}[b]{0.33\textwidth}
        \includegraphics[width=\textwidth]{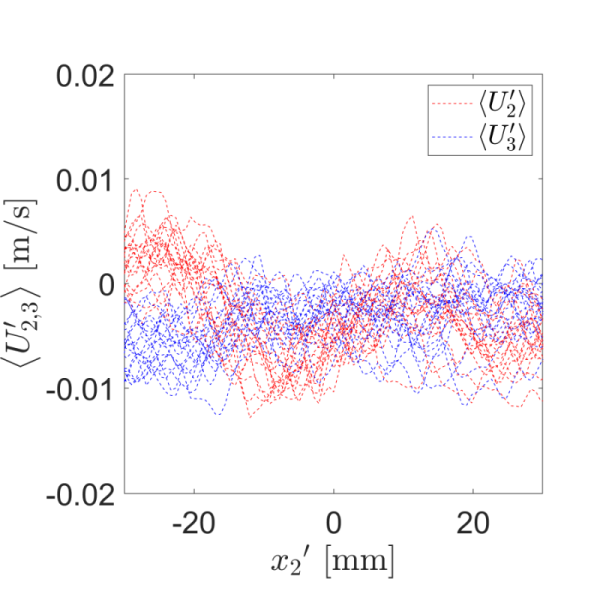} 
          \caption{}
        \label{fig:uncorrected}
        \end{subfigure}
	  \begin{subfigure}[b]{0.33\textwidth}
        \includegraphics[width=\textwidth]{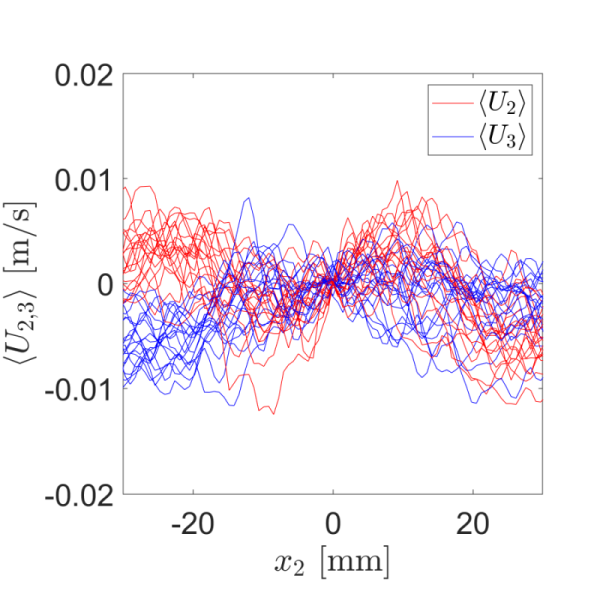} 
          \caption{}
        \label{fig:corrected}
        \end{subfigure}
		\begin{subfigure}[b]{0.37\textwidth}
        \includegraphics[width=\textwidth]{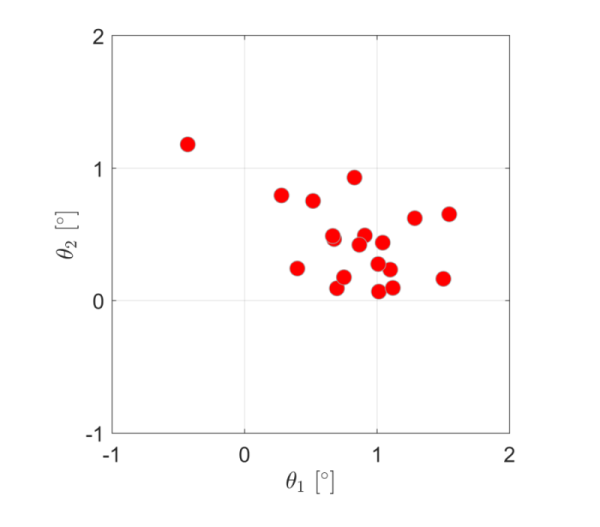} 
          \caption{}
        \label{fig:angles}
        \end{subfigure}

    \caption{a) The radial and out-of-plane mean velocity components in a) laboratory coordinate and b) plume coordinate system averaged across the axial direction, and c) mapping angles ($\theta_1$ and $\theta_2$) based on equation \ref{eq:trans_angle}.}\label{fig:coordinate_transform}
\end{figure}

The measurement bias in $U_2$ and $U_3$ at the plume centerline are captured in figure \ref{fig:uncorrected} that shows the respective mean radial profiles, $\left<U_2^{\prime}\right>$ and $\left<U_3^{\prime}\right>$. We perform this transformation for the 20 replicate datasets independently with correction angles ($\theta_1$, $\theta_2$) reported in figure \ref{fig:angles}.  The bias in the non-zero centerline velocities are reflected in $\theta_1$ and $\theta_2$ which are small and less than 1.5$^\circ$. The two velocity components transformed into the plume coordinate systems are shown in figure \ref{fig:corrected}.  The analysis that follows will be in the plume coordinate system. 


\section{Results and discussion}
\label{sc:results}
\subsection{Mean flow characteristics}
\label{sc:mean}
\begin{figure}
\centering
    \begin{subfigure}[b]{0.75\textwidth}
        \includegraphics[width=\textwidth]{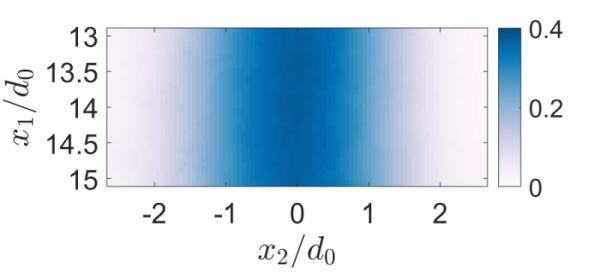} 
          \caption{}
        \label{fig:axial_contour}
        \end{subfigure}
		\begin{subfigure}[b]{0.6\textwidth}
        \includegraphics[width=\textwidth]{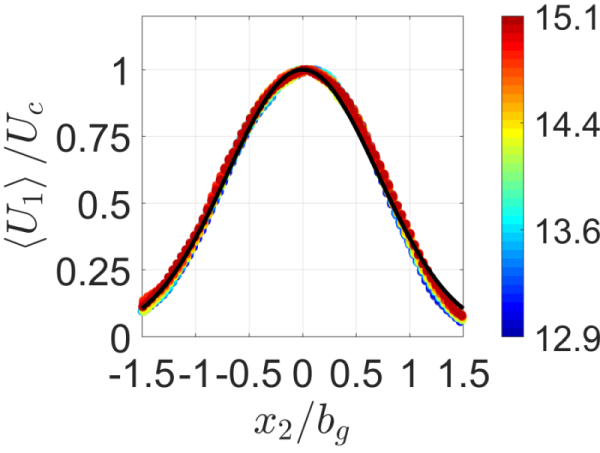} 
          \caption{}
        \label{fig:mean1}
        \end{subfigure}
    \caption{a) 2D intensity field showing the local mean axial velocity  ($\left<U_1\right>$, m/s) across the PIV measurement region, b) radial profiles of non-dimensional $\left<U_1\right>/U_c$ at various axial locations  ($x_1/d_0$) indicated by the colorbar. }\label{fig:mean}
\end{figure}

We first characterize the mean  interstitial fluid  velocity based on approximately 1000 independent PIV snapshots. The mean axial velocity field $\left<U_1(x_1,x_2)\right>$  in dimensional form is shown in figure \ref{fig:axial_contour}. The radial ($x_2$) variations of $\left<U_1\right>$  normalized by the local mean centerline velocity $\left<U_{c}(x_1,0)\right>$ (written as  $U_{c}$ for simplicity from here onward)  are shown in figure \ref{fig:mean1} at representative axial locations ($12.9d_0<x_1<15.1d_0$)  across our measurement region.  The Gaussian plume radius, $b_g$, defined as the $x_2$ location where $\left<U_1\right> = e^{-1}U_c$,  is obtained via a nonlinear least-squares fit of each measured profile to a normalized Gaussian function. The data closely follow the Gaussian curve  (solid line in figure \ref{fig:mean1}),  which is also seen in single- and multi-phase jets and plumes \citep{Darisse2012, Milgram1983, Lai2019}. 

\begin{figure}
\centering
    \begin{subfigure}[b]{0.45\textwidth}
        \includegraphics[width=\textwidth]{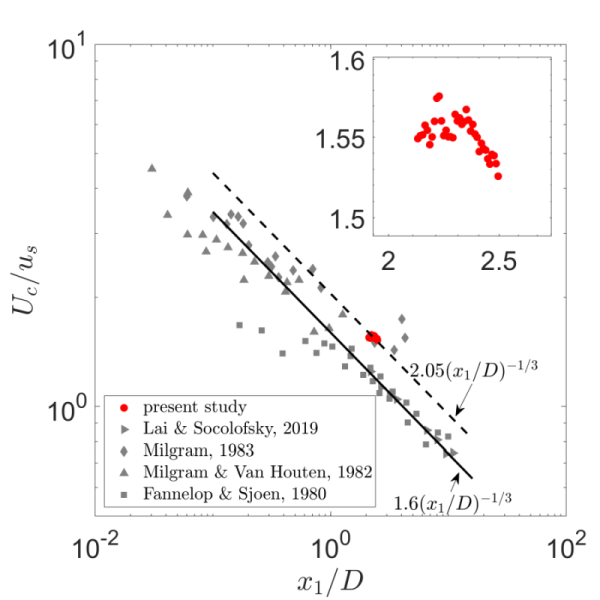} 
          \caption{}
        \label{fig:centerline}
        \end{subfigure}
		\begin{subfigure}[b]{0.45\textwidth}
        \includegraphics[width=\textwidth]{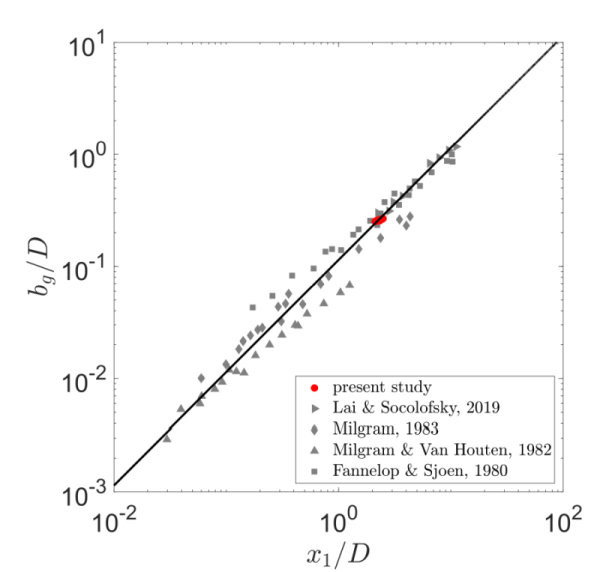} 
          \caption{}
        \label{fig:variation_bg}
        \end{subfigure}
    \caption{Comparison of a) mean centerline velocity, $U_c$ and b) Gaussian plume radius, $b_g$ between a particle plume (current study) and bubble plume (published literature).  }\label{fig:mean_comparison}
\end{figure}

Next, we examine the axial decay of centerline velocity ($U_c$) and the axial growth of Gaussian plume radius ($b_g$) for a particle plume by comparing them to existing data from bubble plumes (see figure \ref{fig:mean_comparison}).  For the purpose of comparison, we use two nondimensional parameters: a velocity scale ($u_s$) and a length scale ($D=gQ_0/4\pi\alpha_0^2u_s^3$). These are based on an integral model of bubble plumes \citep{Bombardelli2007}, in which a constant $\alpha_0 = 0.083$ is used as the entrainment coefficient. The computed entrainment coefficient of our particle plume to be presented later is different from this \emph{a priori} constant. The velocity scale ($u_s$ = 0.23 m/s) is the terminal velocity of a single particle in quiescent fluid and is computed using a simple drag model \citep{Clfit1978}. Based on $Q_0$ = 7.5 $\mathrm{cm^3/s}$, the length scale $D$ is approximately 6.8 cm.  This situates our axial measurement region between 2-2.5$D$.

The overall trends in both centerline velocity and plume radius show similarity between particle and bubble plumes. {The axial variation of $U_c/u_s$ in a particle plume is not measurable from our data, but we can say that it falls above the curve for bubble plumes} \citep{Lai2019} (see figure \ref{fig:centerline}).  The dashed and solid lines in figure \ref{fig:centerline} are the $\mathrm{-1/3^{rd}}$ power-law fits,  with $A$=2.05 (current data, see inset) and $A$=1.6 (\citet{Lai2019}). The variation of $b_g$ for all the data (including ours) is captured by the solid  line ($b_g/D = 0.114x_1/D$) in figure \ref{fig:variation_bg}. Within the limited axial extent ($\approx 0.5D$) of our measurement, the local spreading rate ($\beta = db_g/dx_1$) is difficult to estimate. However, justified based on the collapse of our data on the solid line in figure \ref{fig:variation_bg}, we will use $\beta=0.114$ in the remaining analyses in this paper.

\begin{figure}
\centering
    \begin{subfigure}[b]{0.41\textwidth}
        \includegraphics[width=\textwidth]{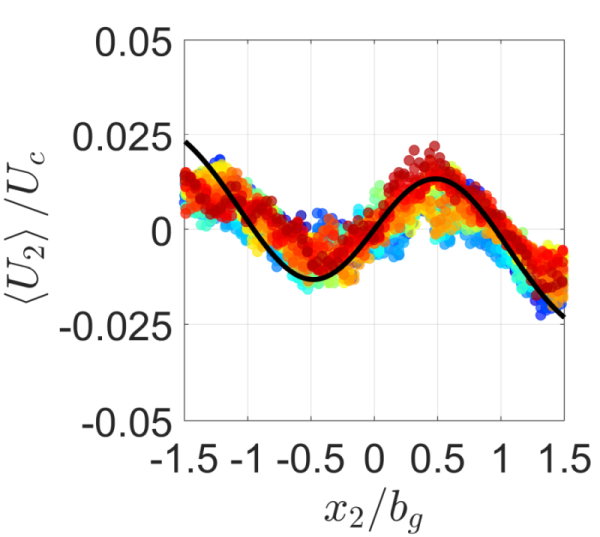} 
          \caption{}
        \label{fig:mean2}
        \end{subfigure}
		\begin{subfigure}[b]{0.49\textwidth}
        \includegraphics[width=\textwidth]{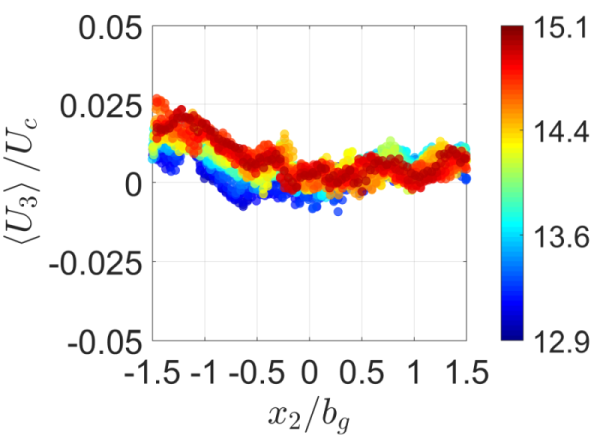} 
          \caption{}
        \label{fig:mean3}
        \end{subfigure}
    \caption{Radial profiles of non-dimensional (a) $\left<U_2\right>/U_c$ and (b) $\left<U_3\right>/U_c$ at various axial locations ($x_1/d_0$) indicated by the colorbar.}\label{fig:meanB}
\end{figure}

Profiles of mean radial ($\left<U_2\right>$) and out-of-plane ($\left<U_3\right>$) velocities at various axial locations are shown in figures \ref{fig:mean2} and \ref{fig:mean3}, respectively. The mean radial velocity exhibits typical jet/plume behavior: within the plume radius ($|x_2/b_g|<1$) the interstitial fluid flows away from the centerline, whereas outside the plume radius ($|x_2/b_g|>1$) the surrounding fluid is entrained into the plume. {The mean out-of-plane velocity should be zero by design (swirling motions are not introduced by the particle feeder)} but it shows some variations and asymmetry far from the plume axis. The solid line in figure \ref{fig:mean2} is a nonlinear least-squares fit to a shape function, 

\begin{equation}
\frac{\left<U_2\right>}{U_{c}}  = -\frac{\alpha}{\eta} \left( (1 - e^{-\eta^2})
- \frac{\beta}{\alpha}\eta^2 e^{-\eta^2}
\right).
\label{eq:radialvel}
\end{equation}
\noindent

\noindent
The shape function in equation \ref{eq:radialvel} is obtained from the radial integration of the continuity equation written in a cylindrical coordinate system after adopting the entrainment hypothesis for jets/plumes and the Gaussian-profile assumption for $\left<U_1\right>$ \citep{Papanicolaou1988,  Lee2003}. Herein, the entrainment coefficient, ${\alpha}$=0.07 is computed as a fitting parameter in equation \ref{eq:radialvel}. The spreading \emph{vs}. entrainment ratio, $\beta/\alpha$, has been measured as 2 for pure jets \citep{Lee2003}, and 1.2 for bubble plumes \citep{Lai2019}. Our results show $\beta/\alpha$ = 0.114/0.07 = 1.63, situating the particle-laden plume between pure jets and bubble plumes. 

\begin{figure}
\centering
        \includegraphics[width=0.5\textwidth]{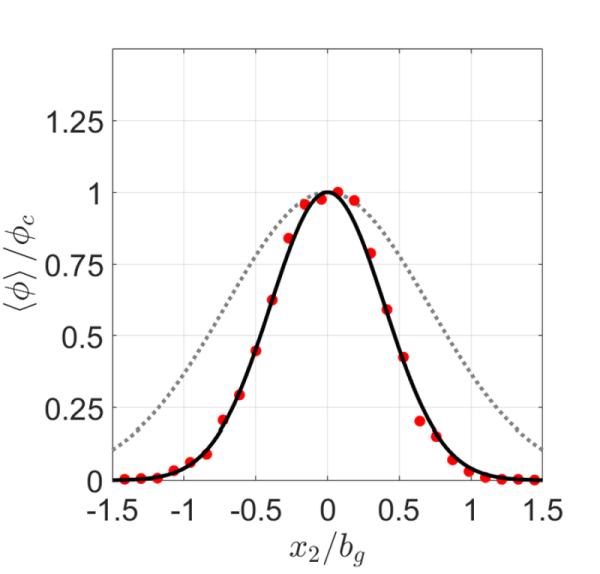} 
    \caption{Normalized distribution of particle number density across the plume averaged along $x_1$.  The solid line is a Gaussian fit to the measured data, and the dotted line is the Gaussian fit to mean axial fluid velocity ($\left<U_1\right>/U_c$) from figure \ref{fig:mean}. The particle half width, $b_{\phi}$ = 0.56$b_g$.}\label{fig:conc}
\end{figure}

{We compute particle number density by counting the particles across each sample (detection method described in section} \ref{sc:particlephase}). The normalized distribution of particle number density $\phi$ across the radius of the plume is shown in figure \ref{fig:conc}. Here, $\left<\phi\right>$ signifies the probability of finding a particle in the specified radial location, and $\phi_c$ is the centerline value, which measures as  0.11 at all axial locations. The distribution of $\phi$ allows us to fit a Gaussian profile and measure the half-width ($b_{\phi}$), {which we use to designate the `particle-core'  from here onward}. The axially averaged {$b_{\phi}$} obtained from the particle concentration profiles, when normalized by the axially averaged $b_g$, shows that $b_{\phi}/b_g$=0.56. This ratio for the bubble plume in \citet{Lai2019} was not measured, but in earlier studies of bubble plumes, {\it e.g.} \citet{Milgram1983}, reported values of $b_{\phi}/b_g$ are in the range of 0.8-0.9. This value is somewhat higher than our value, implying that solid particles spread less rapidly than bubbles. This may be due to the fact that rising bubbles exhibit swirling motions and experience significant lateral lift force when compared to particles \citep{Lai2019}. In addition, large-eddy simulations (LES) for a range of monodispersed and polydispersed bubble plumes in \citet{Fraga2016} {show that the ratio $b_\phi/b_g$ is very sensitive to the size distribution of bubbles across the plume.  They showed that the size distribution across the plume resembles a reverse-Gaussian profile  with larger bubbles populating away from the centerline. This reveals the complexity related to clearly defining a bubble-core  which depends on the distribution of the number-density as well as bubble-size across the plume. A future investigation on bubble plumes examining the relationship between polydispersity and turbulence characteristics will help to better explain the above differences between the particle and bubble plumes.}

\subsection{Fluctuating flow characteristics}
\label{sc:fluc}
\begin{figure}
\centering
    \begin{subfigure}[b]{0.45\textwidth}
        \includegraphics[width=\textwidth]{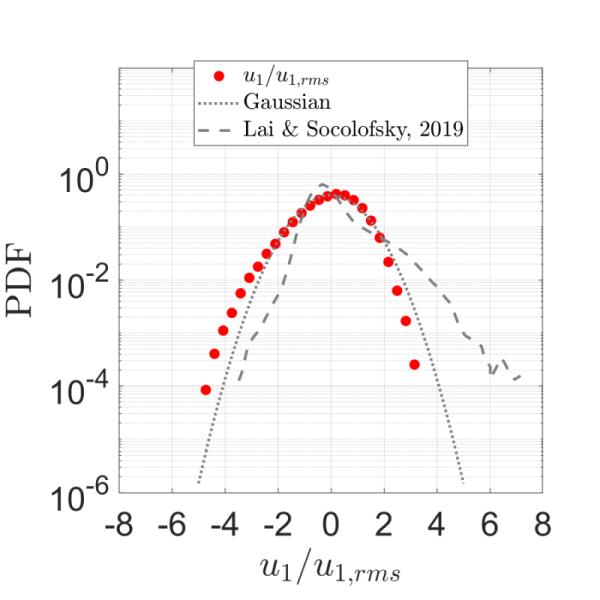} 
          \caption{}
        \label{fig:PDF1}
        \end{subfigure}
		\begin{subfigure}[b]{0.45\textwidth}
        \includegraphics[width=\textwidth]{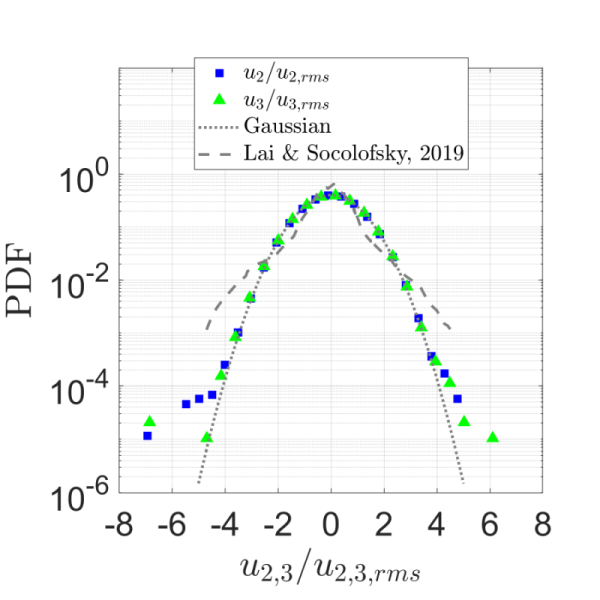} 
          \caption{}
        \label{fig:PDF23}
        \end{subfigure}

    \caption{Comparison of standardized PDF of a) axial: $u_1/u_{1,rms}$, b) radial: $u_2/u_{2,rms}$, and out-of-plane: $u_3/u_{3,rms}$ velocity fluctuations at the centerline of a particle plume (present data) and a bubble plume \citep{Lai2019}. In figure (a), positive values indicate upward-moving fluid in the case of bubble plume and downward-moving fluid in the case of particle plume.}\label{fig:PDF}
\end{figure}

In this section, we examine the nature of the fluctuating components of velocity ($u_i = U_i - \left<U_i\right>$; $u_{i,rms} = \sqrt{\left<u_i^2\right>}$). The normalized probability density functions (PDF) of $u_i$ at the plume centerline is shown in figure \ref{fig:PDF}.  The respective PDF for a bubble plume \citep{Lai2019} is also included in each plot for comparison. Interestingly, the distribution of the axial velocity fluctuations is negatively skewed for a particle plume, such that axial fluctuations  opposite to the direction of particle motion are more common than those along the direction of particle motion. This behavior is strikingly opposite to what is seen in bubble plumes and homogeneous bubble swarms  \citep{Riboux2010, Lai2019, Prakash2016}, for which the axial velocity fluctuations moving in the direction of bubble motion are more common than those moving in the opposite direction. Another way to see this effect is that the mode of the distribution is slightly positive for the particle plume, while it is slightly negative for a bubble plume (see figure \ref{fig:PDF1}). {One implication of this contrasting result is that the fluid flow in a multiphase plume is sensitive to the direction of the plume with respect to gravity; {particle plumes are not a simple reversal of bubble plumes}. } 

The PDFs of radial and out-of-plane velocity fluctuations are symmetric about their means (figure \ref{fig:PDF23}), and closely follow a Gaussian curve. We do not observe the prominent signatures of intermittency in the cross-stream components typically observed in bubble plumes and bubble swarms \citep{Lai2019,Prakash2016}.

\begin{figure}
\centering
    \begin{subfigure}[b]{0.45\textwidth}
        \includegraphics[width=\textwidth]{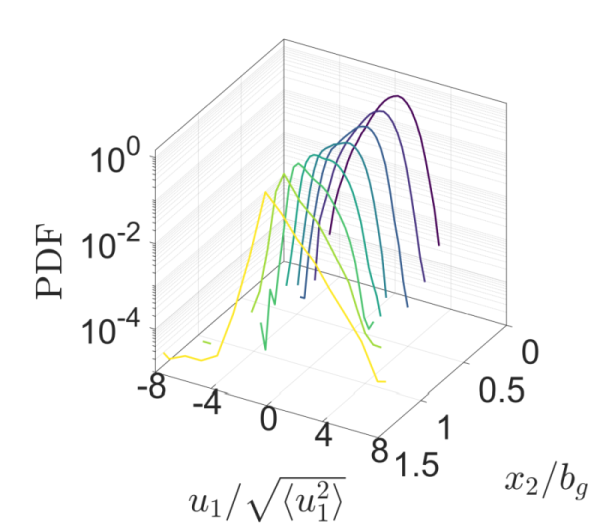} 
          \caption{}
        \label{fig:3dPDF1}
        \end{subfigure}
		\begin{subfigure}[b]{0.5\textwidth}
        \includegraphics[width=\textwidth]{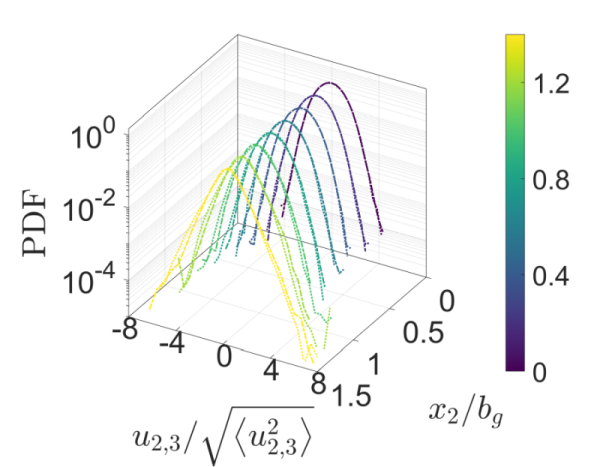} 
          \caption{}
        \label{fig:3dPDF2}
        \end{subfigure}
	
    \caption{Standardized PDF of a) axial: $u_1/u_{1,rms}$, b) radial: $u_2/u_{2,rms}$, and out-of-plane: $u_3/u_{3,rms}$ at various locations across a particle plume. The colorbar indicates the various radial locations ($x_2/b_g$).}\label{fig:3dPDF}
\end{figure}

\begin{figure}
\centering
		\begin{subfigure}[b]{0.3\textwidth}
        \includegraphics[width=\textwidth]{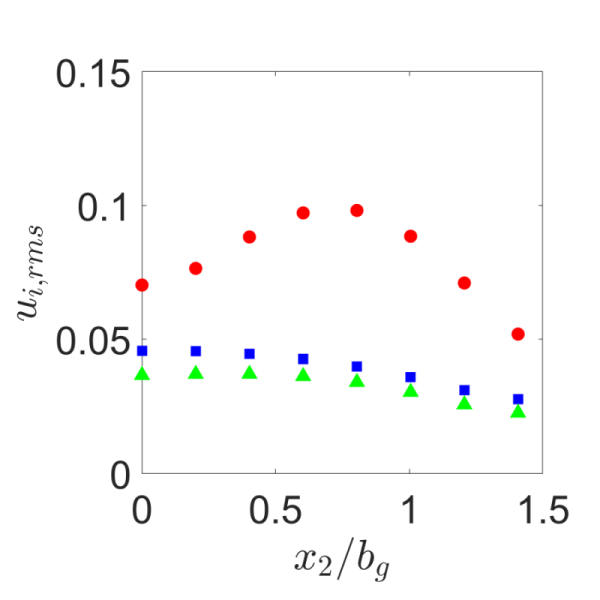} 
          \caption{}
        \label{fig:radstat2}
        \end{subfigure}
	\begin{subfigure}[b]{0.28\textwidth}
        \includegraphics[width=\textwidth]{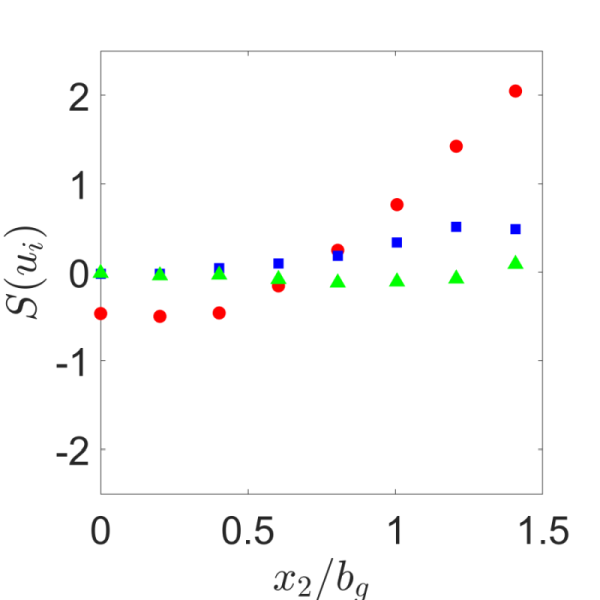} 
          \caption{}
        \label{fig:radstat3}
        \end{subfigure}
\begin{subfigure}[b]{0.28\textwidth}
        \includegraphics[width=\textwidth]{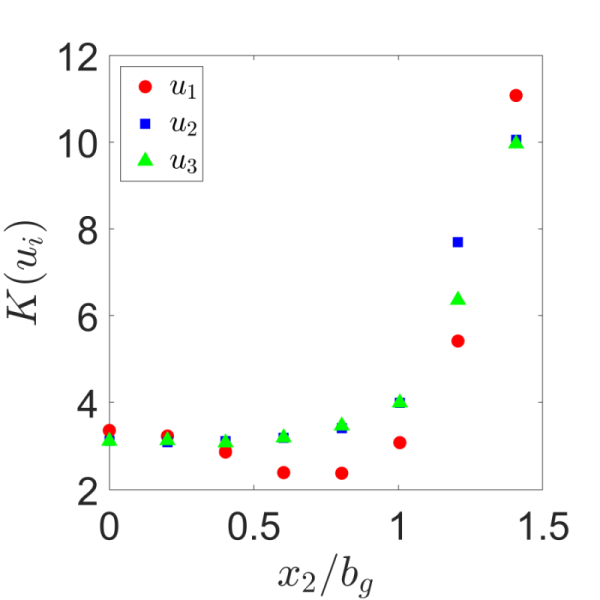} 
          \caption{}
        \label{fig:radstat4}
        \end{subfigure}

    \caption{Higher-order statistics: a) standard deviation, b) skewness and c) kurtosis of the three components of velocity fluctuations across a particle plume.}\label{fig:stats}
\end{figure}

In figure \ref{fig:3dPDF}, we further examine the velocity fluctuations across the plume at different radial locations ($x_2/b_g$). For this purpose, we sort the data into eight equal bins (bin width = 0.16$b_g$) for $x_2\ge 0$ after checking that the choice of bin width does not change the results. Across the plume, the radial ($u_2$) and out-of-plane ($u_3$) velocity fluctuations are symmetric with nearly zero skewness (see figure \ref{fig:3dPDF2}). The axial velocity fluctuations ($u_1$) switch from negatively skewed at the centerline (as discussed earlier) to positively skewed outside the half-radius ($x_2>0.5b_g$) of the plume.  These characteristics are better captured in the statistical moments (rms, skewness and kurtosis) as shown in figure \ref{fig:stats}.  Figure \ref{fig:radstat3} shows that the change of sign in $S(u_1)$ occurs at three-quarters of the plume width $b_g$, which also coincides with the maximum $u_{1,rms}$ (see figure \ref{fig:radstat2}). Both $u_{2,rms}$ and $u_{3,rms}$ show their maximum at the centerline, and are consistently smaller than $u_{1,rms}$.  The radial variation of  kurtosis $K(u_i)$ in  figure \ref{fig:radstat4} captures the increasing flatness of each distribution outside the plume half-radius. 

\subsection{Reynolds stresses}
\begin{figure}
\centering
    \begin{subfigure}[b]{0.46\textwidth}
        \includegraphics[width=\textwidth]{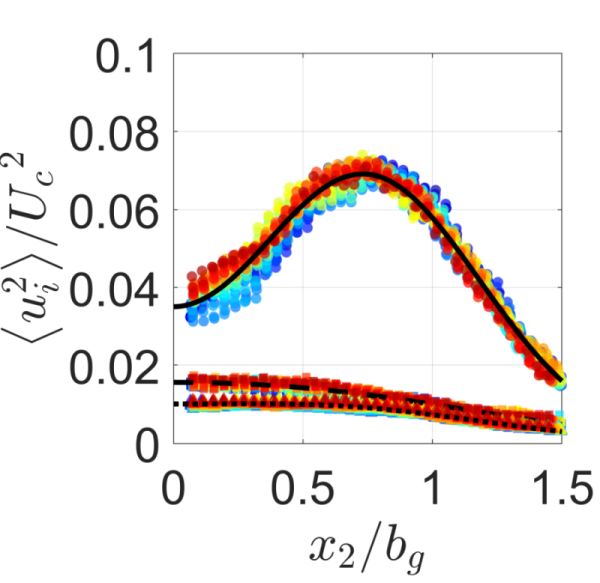} 
          \caption{}
        \label{fig:normal_stress}
        \end{subfigure}
                 \begin{subfigure}[b]{0.43\textwidth}
        \includegraphics[width=\textwidth]{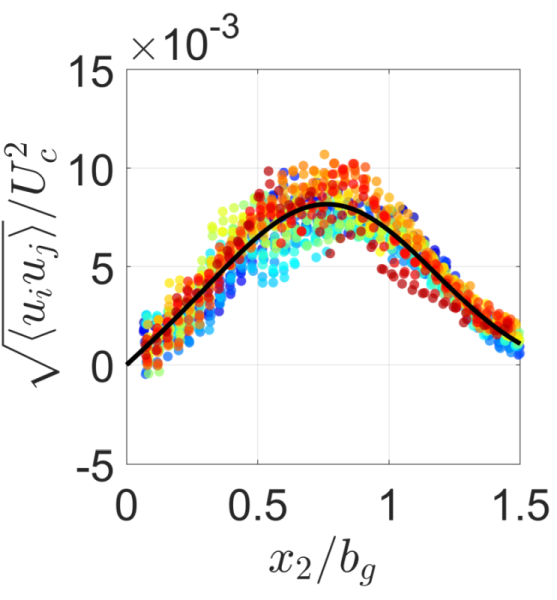} 
          \caption{}
        \label{fig:shear_stress}
        \end{subfigure}
        \begin{subfigure}[b]{0.6\textwidth}
        \includegraphics[width=\textwidth]{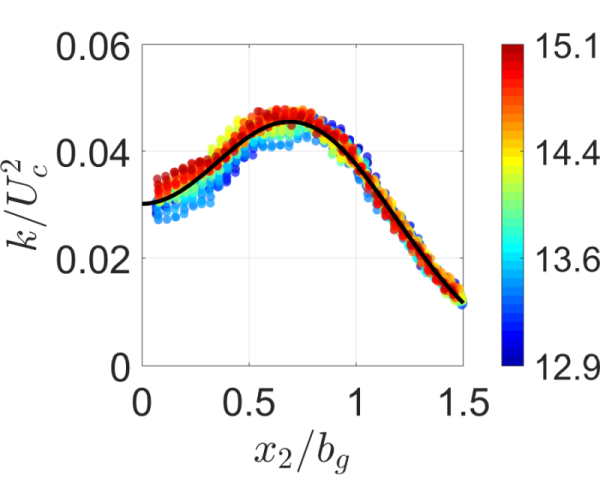} 
          \caption{}
        \label{fig:TKE}
        \end{subfigure}

    \caption{Radial profiles of normalized a) turbulent normal stresses: ${\left<u_1^2\right>}/U_c^2$: $\bigcirc$, ${\left<u_2^2\right>}/U_c^2$: $\Box$, and ${\left<u_3^2\right>}/U_c^2$: $\triangle$, b) normalized turbulent shear stress ${\left<u_1u_2\right>}/U_c^2$, and c) normalized turbulent kinetic energy ($k/U_c^2$) at various axial locations. The colorbars show the normalized axial locations ($x_1/d_0$) indicated by the colorbar.}\label{fig:turb1}
\end{figure}

The Reynolds stresses across the plume are reported at all measured axial locations in figure \ref{fig:normal_stress} (normal stresses) and \ref{fig:shear_stress} (shear stress in the measurement plane). The turbulent kinetic energy ($k$) based on \ref{fig:normal_stress} is shown separately in figure \ref{fig:TKE}.  A nonlinear least-squares fit of the data to a shape-preserving function (see Appendix \ref{appx1}) captures well each profile (see fitted lines in figure \ref{fig:turb1}). Results show that the turbulent kinetic energy is primarily dominated by the axial velocity fluctuations, and that it increases from the plume centerline to its maximum value near the edge of the plume ($x_2 \approx 0.75b_g$). The maximum TKE located at $x_2/b_g=0.75$ is about 44\% of $U_c^2$ and is about 1.5 times the TKE at the plume center. The in-plane shear stress, $\left<u_1u_2\right>/U_c^2$, follows a trend similar to that of a turbulent jet and increases from zero at the centerline to a maximum located near  $x_2/b_g = 0.75$ (see figure \ref{fig:shear_stress}). The location of maximum shear is consistent with that of $\left<u_1^2\right>/U_c^2$, suggesting that the shear stress is dominated by the axial fluctuations $u_1$.

\begin{figure}
\centering
    \begin{subfigure}[b]{0.45\textwidth}
        \includegraphics[width=\textwidth]{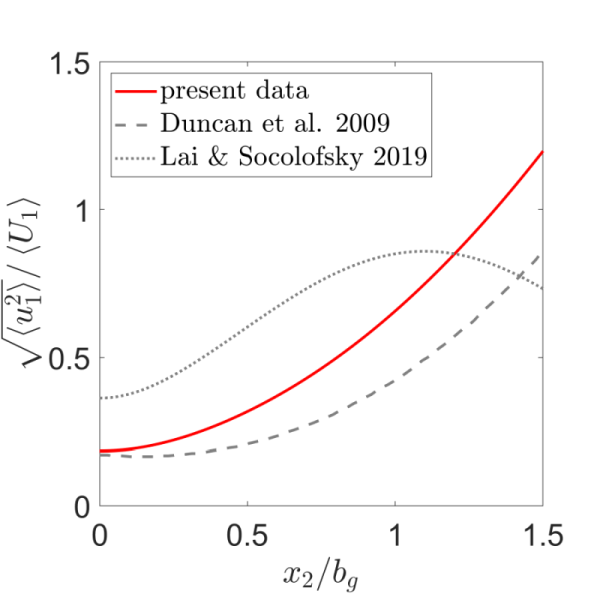} 
          \caption{}
        \label{fig:normal_stress1}
        \end{subfigure}
                 \begin{subfigure}[b]{0.45\textwidth}
        \includegraphics[width=\textwidth]{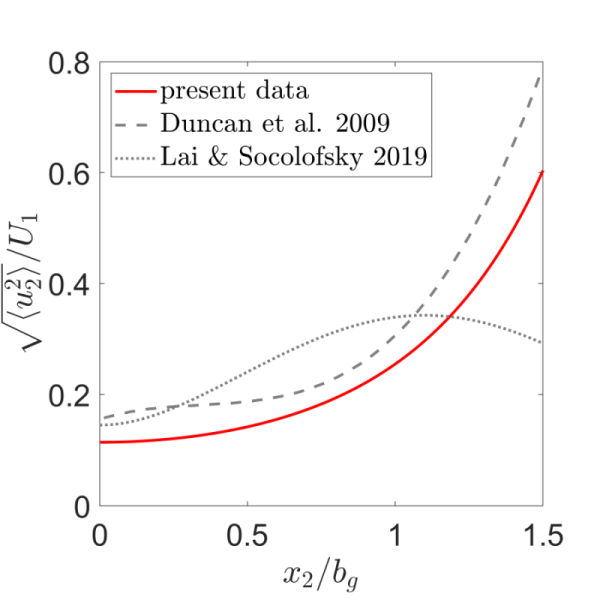} 
          \caption{}
        \label{fig:normal_stress2}
        \end{subfigure}
   \caption{(a) Axial and (b) radial turbulence intensities normalized by the local mean axial velocity $\left<U_1\right>$.}\label{fig:turb1_comp}
\end{figure}

In figure \ref{fig:turb1_comp}, we compare the in-plane turbulent intensities normalized by the local mean axial velocity, ($\sqrt{\left<u_1^2\right>}/\left<U_1\right>$), with two earlier studies on bubble plumes \citep{Duncan2009, Lai2019}. For this comparison, we use the shape functions (solid and dashed lines in figure \ref{fig:normal_stress}.  The present data and the data from \citet{Duncan2009} show reasonably similar trends with unbounded growth away from the centerline {as $\left<U_1\right>$ approaches zero asymptotically outside the plume}. The results from \citet{Lai2019} show somewhat higher turbulent intensity inside the plume core. Also, the growth in their normalized turbulent intensities is not unbounded. \citet{Lai2019} attribute this deviation to finite $\left<U_1\right>$ outside of the plume caused by flow recirculation set up by the tank walls. We do not observe such wall effects in our experiment. The centerline turbulent intensities for the present data are 18\%, 11\% and 9\% for $u_1$, $u_2$, $u_3$, respectively. The axial turbulent intensity ($\sqrt{\left<u_1^2\right>}$) at the centerline for a bubble plume \citep{Lai2019} is also approximately twice the other two normal intensities ($\sqrt{\left<u_{2,3}^2\right>}$), suggesting stronger anisotropy in multiphase plumes when compared to a single-phase jets/plumes in which the ratio $\sqrt{\left<u_{1}^2\right>/\left<u_{2,3}^2\right>}$ is about 1.4 \citep{Wang2002}.

\subsection{Conservation of kinematic momentum flux of the plume}
\begin{figure}
\centering
        \includegraphics[width=0.5\textwidth]{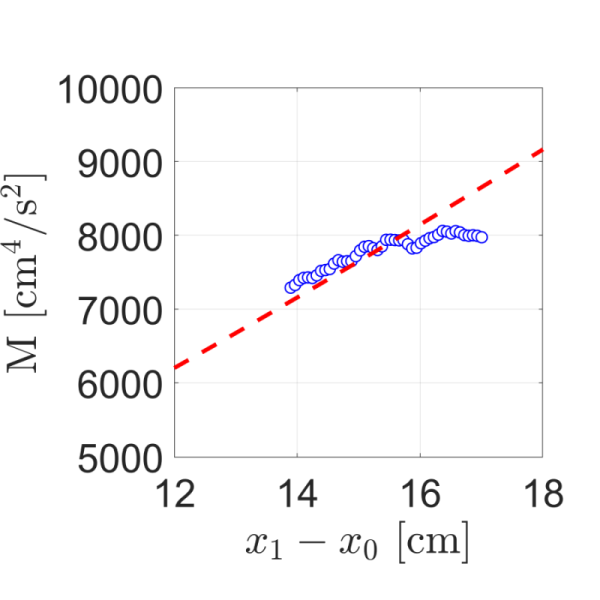} 
        \caption{Variation of total kinematic momentum flux of the induced liquid flow in a particle plume along the axial direction. The dashed line shows a model prediction from equation \ref{eq:mom}.}\label{fig:mom}
\end{figure}
Using the method in \citet{Lai2019}, we assess the conservation of the total kinematic momentum flux, $M=\left<{M}\right>+M_t$, of the induced liquid flow in our particle plume in figure \ref{fig:mom}. Here, the momentum flux contributed by the mean flow is $\left<{M}\right>= 2\pi\int_0^{\infty}x_2U_1^2dx_2$. The momentum flux carried by the turbulence fluctuations is computed as $M_t =  2\pi\int_0^{\infty}x_2\left(\left<u_1^2\right> - 0.5\left(\left<u_2^2\right> + \left<u_3^2\right>\right)\right)dx_2$. The results are compared to the analytical expression for $M(x_1)$ for pure vertical plumes \citep{Lee2003},
\begin{equation}
\centering
M(x_1) = (3\sqrt{2\pi}\beta F_0/4)^{2/3}(x_1-x_{0})^{4/3}.
\label{eq:mom}
\end{equation}

\noindent
Here,  $x_{0}=-5.6 d_0$ is the virtual origin of the plume, and $F_0 = Q_0g$ {is the buoyancy flux  of the particles}. Our results show reasonably good agreement with this model, suggesting that the particle plume obeys the scaling law $M\sim x_1^{4/3}$, which is consistent with buoyancy-driven plumes. The total momentum flux is primarily contributed by the momentum flux due to the mean flow $\left<M\right>$. {The remaining contribution comes from $M_t$, and it is customary to quantify the contribution by the local momentum amplification factor $\gamma = M/\left<{M}\right>$. For our particle plume, $\gamma$ is 1.2, averaged across all axial measurements.} This result shows that $\gamma$ for a particle plume is larger than that for a single-phase jet/plume ($\gamma=1.07 - 1.09$ \citet{Wang2002}) and smaller than a bubble plume ($\gamma = 1.4 - 1.6$, \citet{Lai2019}).  
 
\subsection{Velocity triple-correlation and turbulent transport}
\begin{figure}
\centering
    \begin{subfigure}[b]{0.46\textwidth}
        \includegraphics[width=\textwidth]{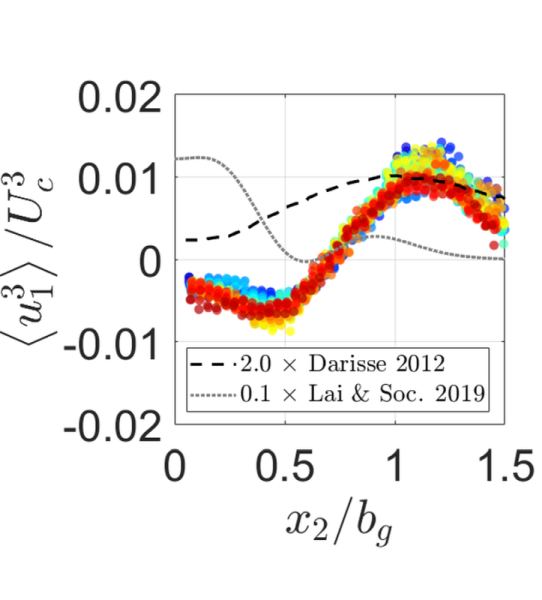} 
          \caption{}
        \label{fig:turb3plea}
        \end{subfigure}
         \begin{subfigure}[b]{0.52\textwidth}
        \includegraphics[width=\textwidth]{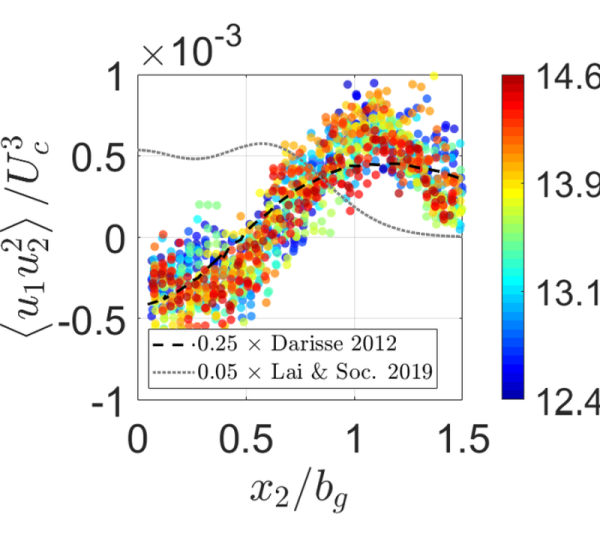} 
          \caption{}
        \label{fig:turb3pleb}
        \end{subfigure}
        
		        \begin{subfigure}[b]{0.4\textwidth}
        \includegraphics[width=\textwidth]{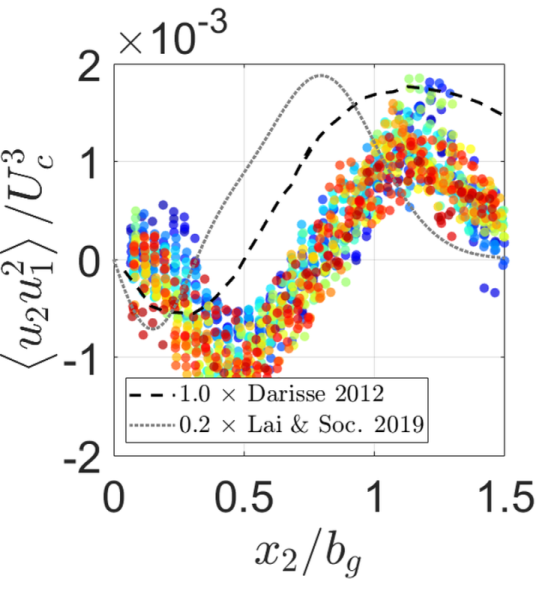} 
          \caption{}
        \label{fig:turb3plec}
        \end{subfigure}
		\begin{subfigure}[b]{0.48\textwidth}
        \includegraphics[width=\textwidth]{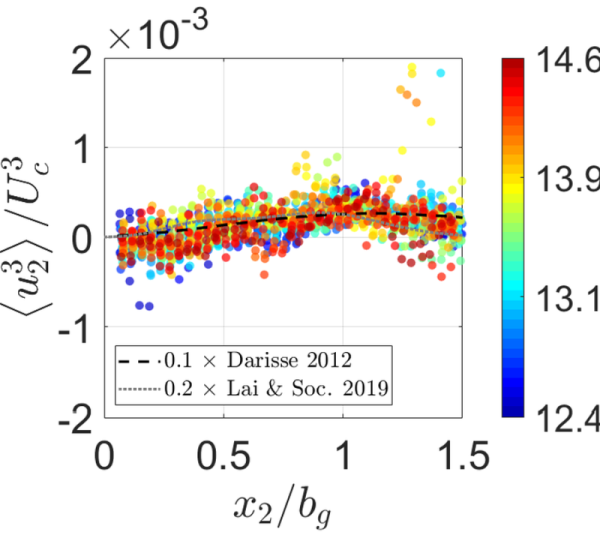} 
          \caption{}
        \label{fig:turb3pled}
        \end{subfigure}
    \caption{Normalized radial profiles of triple correlations (a, b) axial transport ($\left<u_1u_i^2\right>$) and (c, d) radial transport  ($\left<u_2u_i^2\right>$)  of in-plane components of turbulent kinetic energy  in a particle plume at various axial locations ($x_1/d_0$) indicated by the colorbar compared with a single-phase jet \citep{Darisse2012} and a bubble plume \citep{Lai2019}.}\label{fig:turb3ple}
\end{figure}

Because the PDF of axial velocity fluctuations in a particle and a bubble plume are oppositely skewed (see figure \ref{fig:PDF1}), some interesting differences between the two flows can be identified in terms of the velocity triple-correlation terms.  These terms contribute to the transport of turbulent kinetic energy and thus are important in the TKE budget (section \ref{sc:budget}). The radial profiles of the  triple-correlation of in-plane velocity fluctuations normalized by $U_c^3$ are shown  in figure \ref{fig:turb3ple}. To compare the trends we also show the respective profiles for a single-phase jet \citep{Darisse2012} and a bubble plume \citep{Lai2019}. The profiles extracted from the two references are multiplied by a factor for visual comparison  (see legends in figure \ref{fig:turb3ple}).  

Other than $\left<u_1^3\right>$, all triple-correlation profiles for the particle plume show trends similar to a turbulent single-phase jet. The triple-correlation profiles of the first two terms ($\left<u_1^3\right>/U_c^3$ and $\left<u_1u_2^2\right>/U_c^3$) for the bubble plume in \citet{Lai2019}  are significantly different from the single-phase jet and our particle plume, and they show 10--20 times larger magnitude compared to our particle plume results (see figure \ref{fig:turb3plea} and \ref{fig:turb3pleb}). The bubble plume shows a positive $\left<u_1^3\right>$  near the centerline as opposed to the negative $\left<u_1^3\right>$ observed near the core of the particle plume, due to the opposite skewness of their $u_1$ distributions discussed earlier (see figure \ref{fig:PDF1}). 

\begin{figure}
\centering
    \begin{subfigure}[b]{0.49\textwidth}
        \includegraphics[width=\textwidth]{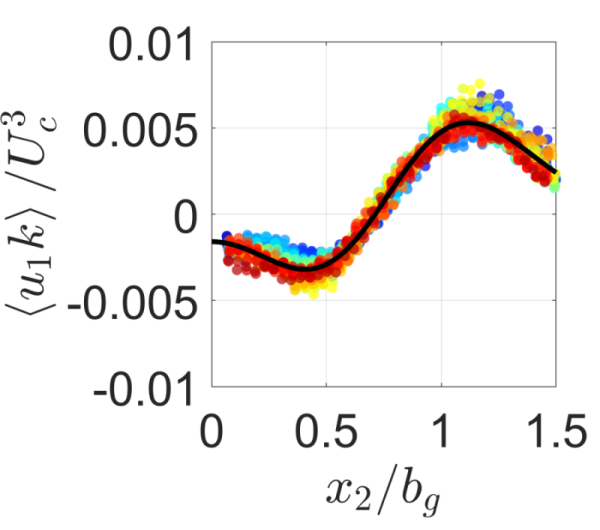} 
          \caption{}
        \label{fig:turb3a}
        \end{subfigure}
		\begin{subfigure}[b]{0.48\textwidth}
        \includegraphics[width=\textwidth]{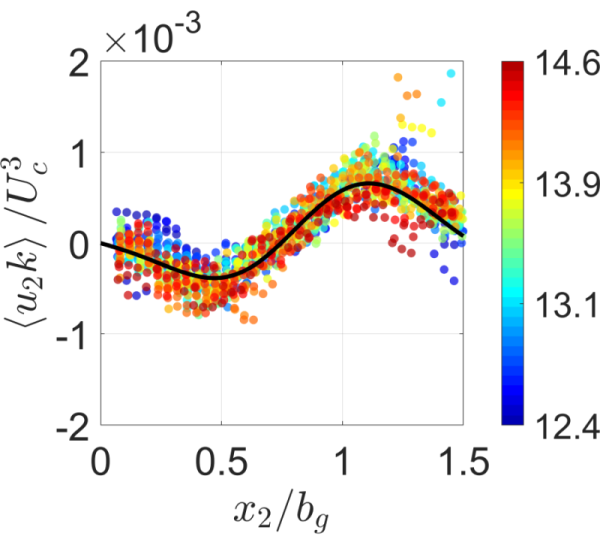} 
          \caption{}
        \label{fig:turb3b}
        \end{subfigure}
	    \caption{Radial profiles of non-dimensional (a) axial ($\left<u_1k\right>/U_c^3$) and (b) radial ($\left<u_2k\right>/U_c^3$) transport of turbulent kinetic energy at various axial locations ($x_1/d_0$) indicated by the colorbar.}\label{fig:turb3}
\end{figure}

Based on the triple-correlation terms computed above, we estimate the axial and radial transport of $k$ in figure \ref{fig:turb3a} and \ref{fig:turb3b}, respectively.  Since the TKE in the particle plume is primarily contributed by the axial stress ($\left<u_1^2\right>$), these two profiles are nearly identical to those in figures \ref{fig:turb3plea} and \ref{fig:turb3plec}. These profiles are fitted to two shape functions (see appendix \ref{appx1}) and are shown as solid lines in figures \ref{fig:turb3a} and \ref{fig:turb3b}, respectively. We use them for estimating the TKE budget in section \ref{sc:budget}.

\subsection{Mean square gradients of velocity fluctuations}
\label{sc:meansqgrad}
\begin{figure}
\centering
    \begin{subfigure}[b]{0.45\textwidth}
        \includegraphics[width=\textwidth]{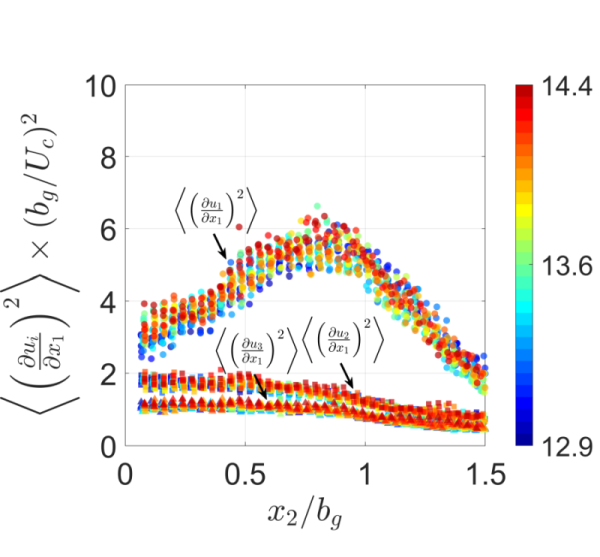} 
          \caption{}
        \label{fig:turb4a}
        \end{subfigure}
		\begin{subfigure}[b]{0.45\textwidth}
        \includegraphics[width=\textwidth]{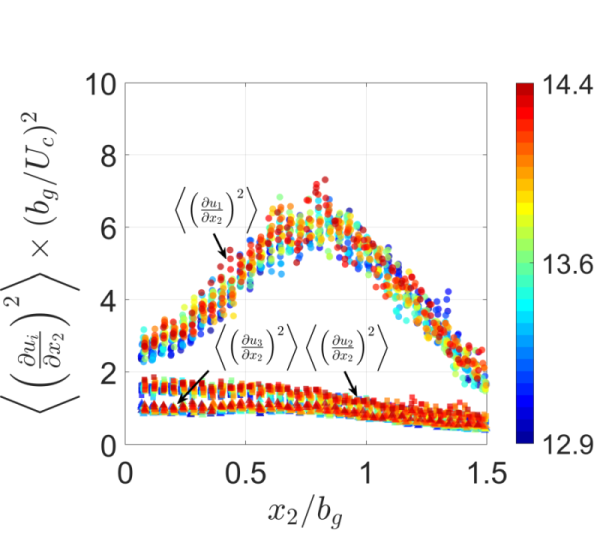} 
          \caption{}
        \label{fig:turb4b}
        \end{subfigure}

    \caption{Radial profiles of non-dimensional (a) axial ($\left<u_1k\right>/U_c^3$) and (b) radial ($\left<u_2k\right>/U_c^3$) transport of turbulent kinetic energy at various axial locations ($x_1/d_0$) indicated by the colorbar.}\label{fig:turb4}
    \end{figure}

To test the conditions of local isotropy and local axisymmetry  \citep{George1991}, we compute the in-plane derivatives of all three components of velocity fluctuations employing {the $\mathrm{2^{nd}}$-order-accurate central-difference scheme}.  Table  \ref{tab:grad} and figure \ref{fig:turb4} show the second moments of axial ($\partial/\partial x_1$)  and radial ($\partial/\partial x_2$) derivatives of all three components of velocity. These moments clearly do not satisfy the conditions of  local isotropy (\emph{e.g.} $ \left<\left(\frac{\partial u_1}{\partial x_1}\right)^2\right> \neq \left<\left(\frac{\partial u_2}{\partial x_2}\right)^2\right>$ and $ \left<\left(\frac{\partial u_1}{\partial x_2}\right)^2\right> \neq 2\left<\left(\frac{\partial u_1}{\partial x_1}\right)^2\right>$) nor of local axisymmetry (\emph{e.g.} $\left<\left(\frac{\partial u_3}{\partial x_1}\right)^2\right> \neq \left<\left(\frac{\partial u_2}{\partial x_1}\right)^2\right>$).  
\begin{table}
  \begin{center}
\def~{\hphantom{0}}
  \setlength{\tabcolsep}{24pt}
  \begin{tabular}{l c c}

${i}$ &          $\left<\left(\frac{\partial u_i}{\partial x_1}\right)^2\right>  (s^{-2})$ &             $\left<\left(\frac{\partial u_i}{\partial x_1}\right)^2\right>  (s^{-2})$  \\ [6pt]
1 & 1555 [1536, 1577] &  1509 [1487, 1530]  \\
2 & 624 [619, 629] & 565 [561, 568.5]  \\
3 & 399 [396, 402] & 368.5 [366, 371]  \\

\end{tabular}
\end{center}
\caption{The second moment of the six available components of the velocity gradient tensor averaged axially and radially. The quantities within the bracket indicate the 95\% uncertainty bounds obtained with the bootstrap method}
\label{tab:grad}
\end{table}

\noindent
The violation of local axisymmetry was also observed for bubble plumes \citep{Lai2019}, suggesting that this is a common characteristic of multiphase plume turbulence {contrary to single-phase jets}.    
    
\begin{figure}
\centering
        \includegraphics[width=0.6\textwidth]{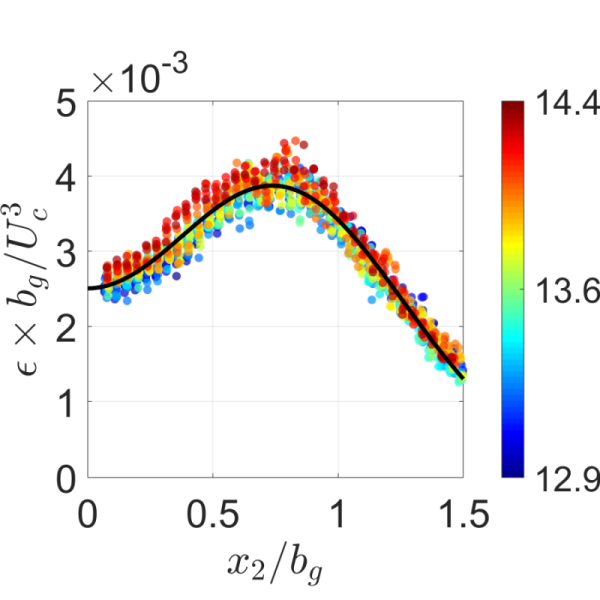} 
           \caption{Radial profiles of non-dimensional mean dissipation rate  at various axial locations ($x_1/d_0$) indicated by the colorbar.}\label{fig:turb5}
    \end{figure}

Computing the true viscous dissipation rate (stress power) requires all nine components of the velocity gradient tensor and each component has to be adequately resolved, which is not easy to achieve experimentally. Instead, we compute the pseudo-dissipation rate $\left<\epsilon\right>$:
\begin{equation}
\begin{aligned}
\left<\epsilon\right> = \nu  & \Biggl[\left<\left(\frac{\partial u_1}{\partial x_1}\right)^2\right>  + \left<\left(\frac{\partial u_1}{\partial x_2}\right)^2\right> + \left<\left(\frac{\partial u_1}{\partial x_3}\right)^2\right> \\
& + \left<\left(\frac{\partial u_2}{\partial x_1}\right)^2\right>  + \left<\left(\frac{\partial u_2}{\partial x_2}\right)^2\right> + \left<\left(\frac{\partial u_2}{\partial x_3}\right)^2\right> \\ 
&+ \left<\left(\frac{\partial u_3}{\partial x_1}\right)^2\right>  + \left<\left(\frac{\partial u_3}{\partial x_2}\right)^2\right> + \left<\left(\frac{\partial u_3}{\partial x_3}\right)^2\right> \Biggr].
\end{aligned}
\label{eq:dissful}
\end{equation} 
\noindent
Here, $\nu = 1.02 \times 10^{-6}$ $\mathrm{m^2/s}$ is the kinematic viscosity of saline water at 23$^\circ$ C.  In our calculations all but the three out-of-plane derivative terms in equation \ref{eq:dissful}  are computed directly from our stereo-PIV  data, and the missing out-of-plane mean-square derivatives ($\left<\left(\frac{\partial u_i}{\partial x_3}\right)^2\right>$) are replaced by $\left<\left(\frac{\partial u_i}{\partial x_2}\right)^2\right>$. The mean pseudo-dissipation rate $\left<\epsilon\right>$ across the plume is shown for various axial locations in figure \ref{fig:turb5}.  

To assess the accuracy in the dissipation calculation, we adopt a method based on conservation of total kinetic energy for multiphase plume in a Lagrangian framework as outlined in \citet{Lai2019}. In essence, this method estimates the balance of the total kinetic energy as production minus dissipation:
\begin{equation}
\centering
\iiint_V \frac{D}{Dt}\left(\left<K\right>+\left<k\right>\right)dV = - f_c\iiint_V \left<\epsilon\right> dV + \iiint_V \frac{\Delta \rho}{\rho} \left<U_1\right>gdV,
\label{eq:cons}
\end{equation}
\noindent
and calculates the correction factor $f_c$ to balance the two sides. The mean kinetic energy, $\left<K\right> \approx \left<U_1\right>^2/2$ because $U_1\gg U_2,U_3$. This equation can be simplified in the cylindrical coordinate system ($r, \theta, z$) by invoking steady state as follows. The left hand side of the equation is computed using the  shape functions of the mean radial profiles for $\left<U_1\right>$ and $\left<k\right>$  (see section \ref{sc:mean} and \ref{sc:fluc}). The first term on the right hand side is obtained by using $\left<\epsilon\right>$ at multiple axial locations. The last term in equation \ref{eq:cons} refers to the kinetic energy production due to the particle buoyancy flux and equals $CF_0\Delta x_1$, where the prefactor $C = \frac{1+\lambda^2}{\pi \lambda^2}\sqrt{4/\gamma}\int_0^{\infty}2\pi\eta e^{-\left(\frac{2\lambda^2+1}{\lambda^2}\right)\eta^2}d\eta$ depends on $\lambda = \left<b_{\phi}\right>/\left<b_{g}\right>$ and the momentum amplification factor $\gamma$. Based on $\lambda = 0.6$ and $\gamma = 1.2$, we obtain $C=1.44$. The simplified equation for a volume bounded by two axial locations $x_1$ and $x_2$ is written as:

\begin{equation}
\centering
\left[\left(0.5I_1 + I_2\right)\left(b_g^2U_c^3\right)\right]_{x_1}^{x_2} = f_c\int_{z_1}^{z_2}I_3b_g{U_c^3}dx + 1.44F_0\Delta x_1
\label{eq:cons2}
\end{equation}
\noindent
Here, $I_1$, $I_2$ and $I_3$ are the axisymmetric surface integrals ($2\pi\int_{0}^{\infty} \left( . \right) r dr$) of  the respective terms inside the volume integral in equation  \ref{eq:cons} (summarized in Table \ref{tab3}). 

\begin{table}
  \begin{center}
\def~{\hphantom{0}}
  \setlength{\tabcolsep}{4pt}
  \begin{tabular}{l c c c}
  Term in eq. \ref{eq:cons} & Full form & Similarity form & Value \\ \hline 
   Flux of $K$, $I_1$ & $2\pi\int_0^{\infty}\left<U_z\right>\left<K\right>rdr$ & $2\pi\int_0^{\infty}\eta e^{-3\eta^2}d\eta$ & 1.047\\
   Flux of $k$, $I_2$ & $2\pi\int_0^{\infty}\left<U_z\right>\left<k\right>rdr$ & $2\pi\int_0^{\infty}\eta e^{-\eta^2}F_4(\eta)d\eta$  & 0.095 -- 0.105\\
  Dissipation, $I_3$ & $2\pi\int_0^{\infty}\left<\epsilon\right>rdr$ & $2\pi\int_0^{\infty}\eta F_5(\eta)d\eta$  & 0.021 -- 0.027 \\ \hline
  $K \approx U_1^2/2$ & $k = (u_1^2 + u_2^2 + u_3^2)/2$ & & 
  \end{tabular}
  \end{center}
  \caption{Simplified similarity expressions and values of various integrals used in equation \ref{eq:cons2}. The shape functions $F_4(\eta)$ and $F_5(\eta)$ are summarized in Appendix \ref{appx1}.}
  \label{tab3}
  \end{table}

We test the balance of equation \ref{eq:cons2} at multiple axial locations separated by $1.5b_g$ and have found $f_c$ to vary between 2.8 and 5.1 with a mean value of 4.0. This method estimates the amount by which $\left<\epsilon\right>$ is underestimated by our evaluation of equation \ref{eq:budget}, but it is unable to correct $\left<\epsilon\right>$ locally along the radius of the plume. Because of this limitation in our measurement of $\left<\epsilon\right>$, we use the corrected dissipation profile only for a qualitative assessment of the TKE budget discussed in the following section.

\subsection{Turbulent kinetic energy budget}
\label{sc:budget}
In the cylindrical coordinate system, the time-averaged transport equation for $k$ of an axisymmetric, steady, turbulent plume without swirl is \citep{Kataoka1989}:

\begin{align}
0 &= \overbrace{-\left(\left<U_r\right>\frac{\partial k}{\partial r} + \left<U_z\right>\frac{\partial k}{\partial z}\right)}^{\emph{A}} \nonumber \\
  & \qquad{} \overbrace{ -\left(\frac{1}{r}\frac{\partial \left<{u_rk}\right>}{\partial r} + \frac{\partial \left<{u_zk}\right>}{\partial z}\right)}^{\emph{T}} \nonumber \\
  & \qquad{} \overbrace{ -\frac{1}{\rho}\left(\frac{1}{r}\frac{\partial \left<{u_rp}\right>}{\partial r} + \frac{\partial \left<{u_zp}\right>}{\partial z}\right)}^{\text{$T_p$}} \nonumber \\
  & \qquad{} \overbrace{-\left(\left<{u_z^2}\right>\frac{\partial \left<U_z\right>}{\partial z}+\left<{u_r^2}\right>\frac{\partial \left<U_r\right>}{\partial r} + \frac{\left<U_r\right>\left<{u_\theta^2}\right>}{r} + \left<{u_zu_r}\right>\left( \frac{\partial \left<U_z\right>}{\partial r} +  \frac{\partial \left<U_r\right>}{\partial z} \right)\right)}^{\text{$P_s$}} \nonumber \\
  & \qquad{} - \left<\epsilon\right> \nonumber \\
  & \qquad{} + P_b
\label{eq:budget}
\end{align}
\noindent

In addition to the terms already present in single-phase jets/plumes ($A$: advection, $P_s$: shear-production, $T$: turbulent transport, and  $T_p$: pressure transport), a multiphase plume has another source term ($P_b$) that represents the interfacial energy transfer at the boundaries between the dispersed phase and the surrounding fluid. Obtaining the complete TKE budget for a particle plume is challenging for multiple reasons. First, it is difficult to accurately measure experimentally the mean dissipation rate $\left<\epsilon\right>$ because of the under-resolved velocity gradients (as seen in section \ref{sc:meansqgrad}). Second, there is no existing model for pressure-velocity correlation in multiphase flow turbulence, and therefore the pressure transport term ($T_p$) requires an assumption. Finally, the interfacial energy transfer ($P_b$) is the most complex and inaccessible term, and it requires the local distribution of particles and their relative velocity with the surrounding fluid \citep{Santarelli2016}. In an ideal scenario with all other terms measured correctly, the term $P_b$ can be sought as the closing term of equation \ref{eq:budget}, such that $P_b = \left<\epsilon\right>-(A+P_s+T+T_p)$. We will use this method to estimate $P_b$ from our data. 

\begin{figure}
    \centering
\begin{subfigure}[b]{0.45\textwidth}
        \includegraphics[width=\textwidth]{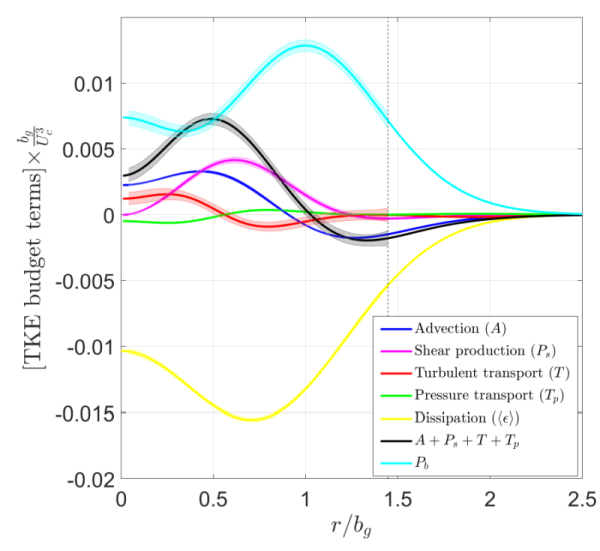} 
          \caption{}
        \label{fig:budget1}
        \end{subfigure}
		\begin{subfigure}[b]{0.48\textwidth}
        \includegraphics[width=\textwidth]{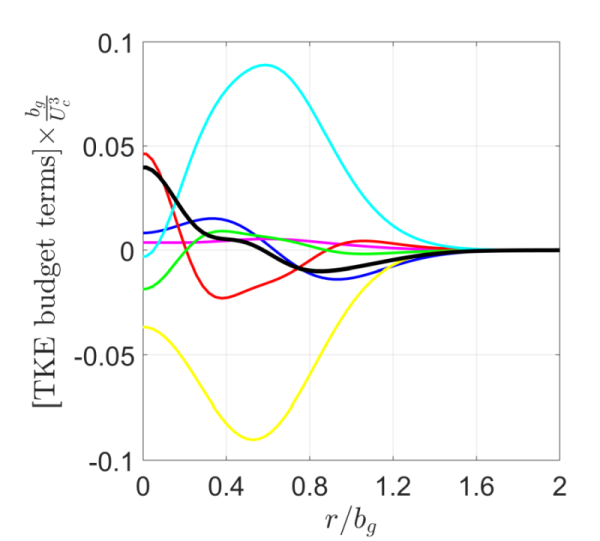} 
          \caption{}
        \label{fig:budget2}
        \end{subfigure}
        \caption{Various terms in the turbulent kinetic energy budget for a a) particle plume and b) bubble plume reproduced from \citet{Lai2019}. {The error band in each term (shown upto the extent of measurement window) represents uncertainty in computing the term.}}
        \label{fig:budget}
\end{figure}

We directly compute $A$, $P_s$ and $T$ from our experimental data. The expressions for these in terms of similarity variables are summarized in Appendix \ref{appx2}. {Each plot contains an error-band corresponding to the 95\% confidence interval in the constituent fits. Each uncertainty band is computed via the rule of uncertainty propagation described in} \citet{Charonko2017}. The pressure transport term ($T_p$), not directly available from experiment, is replaced by Lumley's model $T_p = -2/5T$, widely used for single-phase turbulence \citep{Lumley1978}. These four terms and the mean dissipation rate ($\left<\epsilon\right>$)  multiplied by the correction factor $f_c = 4$ are shown in figure \ref{fig:budget1}. The remaining terms of the balance equation are the interfacial energy transfer ($P_b$) from particles to fluid and out-of-balance experimental error (OOBE).  The budget terms for a bubble plume from \cite{Lai2019} are reproduced in figure \ref{fig:budget2} for comparison. Each budget term is normalized by $U_c^3/b_g$; the order-of-magnitude difference in the two budgets arises because of the difference in the downstream measurement locations with respect to the source, leading to the differences in $U_c$ and $b_g$ (see figures \ref{fig:centerline} and \ref{fig:variation_bg}). Below we discuss the comparison of the TKE budget for the two {multiphase} plumes in light of previous experiments on a single-phase {jet} \citep{Lai2018} and a variable-density single-phase (VDSP) jet \citep{Charonko2017}.  The qualitative summary of this comparison is provided in table \ref{tab4}.

 The shape of the advection ($A$) profiles are {very similar} across all four plumes with a positive central lobe inside the plume core. The shear production term ($P_s$) in the particle plume increases from zero at the centerline to a dominant peak {at about $r/b_g = 0.65$ near the edge of the plume}. This behavior strongly resembles a single-phase jet both in shape and relative magnitude. The $P_s$ profile in the VDSP jet also matches this trend, except that a negative $P_s$ is observed at the jet centerline. By contrast, the bubble plume shows almost negligible $P_s$ across the radius of the plume. The turbulent transport term ($T$) in the particle plume exhibits a positive peak near the centerline and a negative lobe near the edge of the plume. The bubble plume and the VDSP jet are similar, but with a larger peak in the former. The single-phase {jet} does not match the others, \emph{i.e.} near the centerline the turbulent transport is consistently observed to be negative or marginally positive \citep{Darisse2015}.

\begin{table}
 \begin{center}
\def~{\hphantom{0}}
 \setlength{\tabcolsep}{4pt}
 \begin{tabular}{l c c c c}
 TKE budget term  & \multicolumn{2}{c}{Multiphase}  & \multicolumn{2}{c}{Single phase}  \\ \hline
                  & Particle  & Bubble  & Jet & VD jet \\
Advection, $A$  &   \textcolor{green}{\cmark} &   \textcolor{green}{\cmark}  &  \textcolor{green}{\cmark}  &    \textcolor{green}{\cmark}\\
Shear production, $P_s$  &   \textcolor{green}{\cmark}&   \textcolor{red}{\xmark}   &     \textcolor{green}{\cmark}  &   \textcolor{red}{\xmark} \\
Turbulent transport, $T$  &   \textcolor{green}{\cmark}&  \textcolor{green}{\cmark}    &     \textcolor{red}{\xmark}    &    \textcolor{green}{\cmark}     \\ \hline    
\textcolor{green}{\cmark}: similar profile shapes &\multicolumn{2}{c}{\textcolor{red}{\xmark}: dissimilar profile shapes}  & &  \\
 \end{tabular}
 \end{center}
 \caption{Qualitative comparison summary of three TKE budget terms in our particle plume and those in a bubble plume \citep{Lai2019}, a single-phase {jet} \citep{Darisse2015, Lai2018}, and a variable density single-phase (VDSP) jet \citep{Charonko2017}. }
 \label{tab4}
 \end{table}

 Because the  energy production at the two-phase boundaries ($P_b$) is computed indirectly, the peak location in $P_b$ is sensitive to the experimental error in $\left<\epsilon\right>$. Despite this limitation, it is worth noting  that in both particle and bubble plumes, the sum ($A+T+P_s+T_p$) is significantly smaller than the mean dissipation rate outside the plume core ($r>0.5b_g$). This results in the TKE budget being an approximate balance between $P_b$ and $\left<\epsilon\right>$.  

\subsection{Two-point correlation and energy spectra}

\begin{figure}
\centering
        \includegraphics[width=0.5\textwidth]{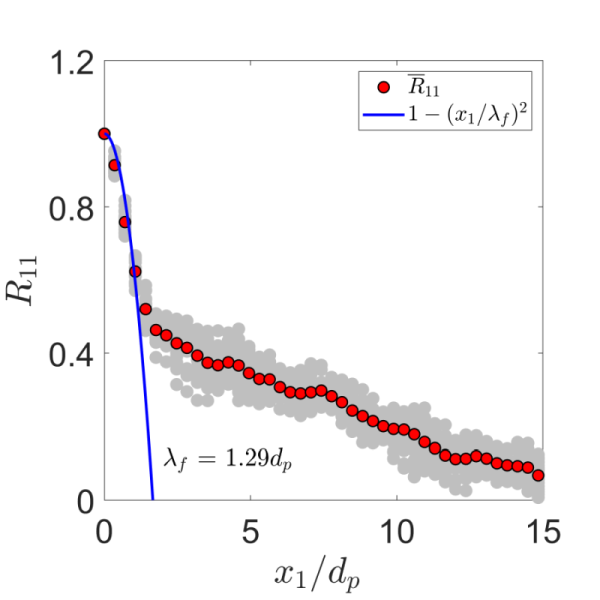} 
    \caption{Spatial autocorrelation $R_{11}$ along longitudinal (axial) direction of the flow. The shaded circles show autocorrelation at various radial location ($x_2/b_g<0.5$), the filled circles are their mean. The solid line is an osculating parabola fit to the first four data points in $\bar{R}_{11}$.}\label{fig:R11}
    \end{figure}

We  obtain the two-point statistics from the spatial autocorrelation function $R_{11}$ computed along the axial measurements of axial velocity fluctuations, $u_1$. Figure \ref{fig:R11} shows $R_{11}$ inside the plume core ($x_2/b_g<0.5$) at various radial locations (shaded circles) and their mean $\bar{R}_{11}$ (red circles). The horizontal axis is normalized by the particle diameter $d_p$. We estimate the Taylor microscale ($\lambda_f$) of this flow by fitting an osculating parabola, $1-\frac{x_1^2}{\lambda_f^2}$  \citep{Pope2000} to the first four points of $\bar{R}_{11}$ (see solid line in figure \ref{fig:R11}), yielding $\lambda_f = 1.3d_p$, which is of the order of the particle size. The Taylor microscale ($\lambda_f$) is related to the mean square velocity gradient in homogeneous isotropic turbulence \citep{Pope2000} and we use this relation to approximate
\begin{equation}
\centering
\left<\left(\frac{\partial u_1}{\partial x_1}\right)^2\right> \approx \frac{2\left<u_1^2\right>}{\lambda_f^2}.
\label{eq:taylor}
\end{equation}
\noindent
Based on this method, $\left<\left(\frac{\partial u_1}{\partial x_1}\right)^2\right>\left(b_g/U_c \right)^2$ at $x_2=0$  is 3.2$\pm$0.6 and at $x_2 = 0.5b_g$ is 5.8 $\pm$0.6. Within the uncertainty bounds, these values match those obtained from direct computation of gradients in section \ref{sc:meansqgrad} (see figure \ref{fig:turb4a}). 

\begin{figure}
\centering
 \begin{subfigure}[b]{0.48\textwidth}
        \includegraphics[width=\textwidth]{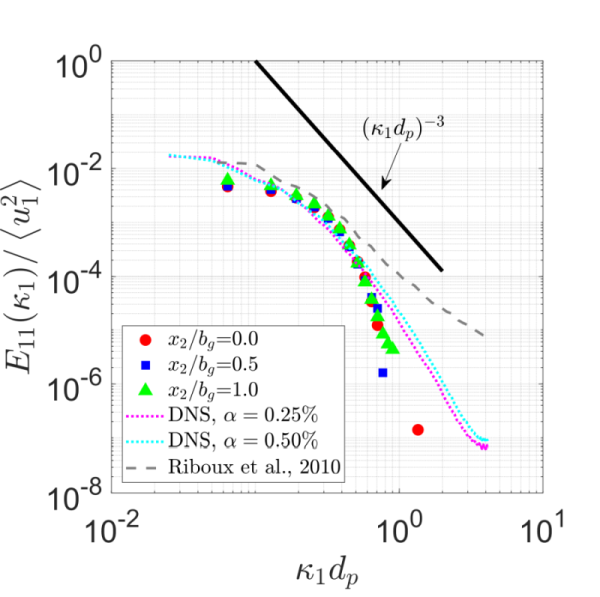} 
          \caption{}
       \label{fig:spectra1}
       \end{subfigure}
 \begin{subfigure}[b]{0.48\textwidth}
        \includegraphics[width=\textwidth]{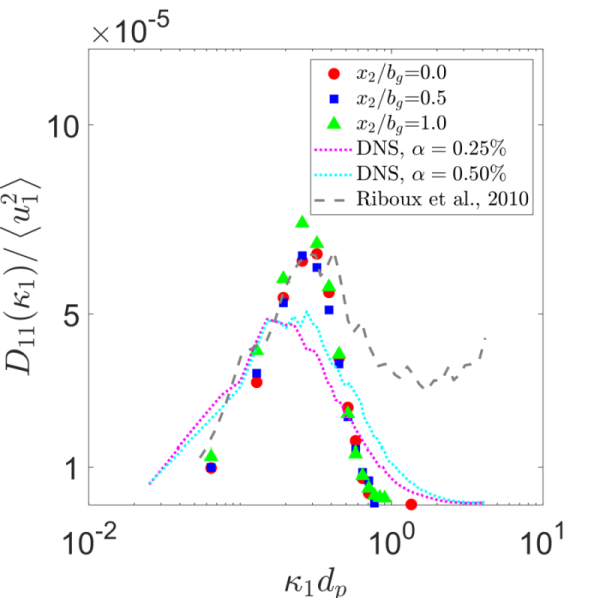} 
          \caption{}
       \label{fig:spectra2}
       \end{subfigure}
    \caption{Comparison of a) one-dimensional power spectra $E_{11}(\kappa_1)$ of axial velocity fluctuations and b) one-dimensional dissipation spectra in particle plume with earlier studies related to homogeneous bubble swarms (experiment: \citep{Riboux2010}, DNS simulation: \citet{Lai2018a}).  The horizontal axis is normalized by the wave number corresponding to respective particle diameter ($d_p$). }\label{fig:spectra}
    \end{figure}

\citet{Risso2018} suggests that the inter-scale energy transfer in bubble-induced agitation (BIA) is quite different from the classical picture of shear-induced turbulence (SIT). As a first step toward understanding the energy cascade in particle-plume turbulence, we generate the one-dimensional energy spectrum ($E_{11}$) from the autocorrelation function $\bar{R}_{11}$ \citep{Pope2000}. The normalized spectra for our particle plume is compared with experimental results \citep{Riboux2010} and DNS simulations \citep{Lai2018a} in figure \ref{fig:spectra1}. The result shows striking similarity in the spectra between homogeneous bubble swarms (both experiment and DNS simulation) and our particle plume, all following the earlier prediction of a $\kappa^{-3}$ power law for wave numbers smaller than $1/\lambda_f$. Beyond $\kappa_1 >1/\lambda_f$, both experimental results (ours and \citet{Riboux2010}) show different slopes than the DNS results, which can be attributed to experimental uncertainty in resolving velocity fluctuations below $\lambda_f$. Nonetheless, this result is valuable and it suggests power-law universality of multiphase flow turbulence which is quite different from that for single-phase turbulent shear flows.

The one-dimensional dissipation spectra $D_{11}(\kappa_1) = 2\nu\kappa^2E_{11}(\kappa_1)$ is shown in figure \ref{fig:spectra2}. Quite strikingly, the peaks of dissipation for our data and bubble plume/swarms reside at $\kappa_1d_p \approx 0.2-0.4$. This indicates that in both bubble-- and particle--laden turbulence, production and dissipation occur near the scale of particle size. This lack of scale separation was first postulated for homogeneous bubbly flows in \citet{Lance1991} and is observed in direct numerical simulations reported in \citet{Lai2018a}.

\section{Conclusions}\label{sc:conclusions}
We report an experimental study characterizing the turbulence {inside} a heavy particle plume descending under gravity within a salt-water solution. We measure the three components (2D3C) of the interstitial fluid velocity and the spatial distribution of particles in the central plane of the plume using refractive-index-matched stereoscopic particle image velocimetry. Below we summarize our key findings primarily in the light of the results for a bubble plume \citep{Lai2019}.

The induced liquid flow  inside the plume evolves with the mean flow characteristics of a bubble plume \citep{Lai2019}, that show Gaussian mean axial velocity ($\left<U_1\right>$) profile. The radial profiles of mean particle number density are also Gaussian with a half-width $b_\phi$ equals to 0.56$b_g$, where $b_g$ is the Gaussian half-width of the mean streamwise velocity profile in the fluid.

The turbulence inside the plume is highly anisotropic with maximum streamwise turbulence intensity measuring up to {2.7} times that of the other two components, a result consistent with that observed in a bubble plume.   The PDF of the axial velocity fluctuation ($u_1$) at the centerline of the plume is skewed so that the strongest events are in the direction opposite to mean flow. This behavior is strikingly opposite to that observed previously for a bubble plume and a homogeneous swarm of bubbles.  The turbulent kinetic energy (TKE) and the in-plane shear-stress peak near $\approx 0.75 b_g$, locating the shear layer slightly outside the edge of the particle-core ($b_\phi$).  Although these quantities for a bubble plume in \citet{Lai2019} peaked at a similar location ($\approx 0.7 b_g$),  the location of the shear layer with respect to their bubble-core ($b_\phi$) remains unknown.

Another distinction between the particle and the bubble plume is observed in the fluid shear production term ($P_s$) in the TKE budget. While the shear production in the bubble plume is negligible across the plume compared to the other terms, we observe a distinct peak in $P_s$ at the shear layer region of a particle plume, resembling a single-phase jet. 

The other TKE terms including advection ($A$) and the turbulent transport ($T$) in the two flows exhibit similar profiles. Further, we show that the turbulent transport ($T$) term follows similar qualitative profiles for both plumes and also an earlier reported variable-density single-phase jet \citep{Charonko2017}, all of which differ from a typical single-phase jet.  The difference in the relative magnitude in $T$ at the centerline between the two plumes is attributed to the difference in their third moment (skewness) of the axial velocity fluctuations.


Despite the above differences inside the core, both particle and bubble plumes show qualitatively that the TKE production by the particles ($P_b$) approximately balances the mean dissipation rate ($\left<\epsilon\right>$) away from the centerline. Further, the one-dimensional spectrum in the particle plume exhibits the $\mathrm{-3^{rd}}$ power-law consistent with bubble plume and homogeneous swarms of bubbles. {These two results support the notion that there is a lack of separation between the scales of production and dissipation in multiphase turbulent flows like ours.}

\section*{Acknowledgement}
This research was supported by the National Science Foundation (Grant No. OCE-
1334788) and by The Gulf of Mexico Research Initiative. The authors also thank Michael Heath for his help in the design of the experimental setup and data collection.
\appendix
\section{}
\label{appx1}
The three components of normal stresses, the turbulent kinetic energy ($k$), and the mean dissipation rate $\left<\epsilon\right>$ are fitted into a shape preserving double-Gaussian curve, given in equation \ref{Apeq:1}. All terms are expressed in similarity variable $\eta = r/b_g$.

\begin{equation}
\centering
F_i(\eta) = f_1\left[\mathrm{exp}\left({-\left(\frac{\eta - f_3}{f_2}\right)^2}\right) + \mathrm{exp}\left({-\left(\frac{\eta + f_3}{f_2}\right)^2}\right)   \right]
\label{Apeq:1}
\end{equation}

\begin{table}[H]
  \begin{center}
\def~{\hphantom{0}}
  \setlength{\tabcolsep}{12pt}
  \begin{tabular}{c c c c c}
  Function ($F_i(\eta)$)  & Parameter & $f_1$ &  $f_2$ &  $f_3$ \\\\
  $F_1$ &$\left<u_z^2\right>/U_c^2$ &   0.069 &   0.74 & 0.63 \\
  $F_2$ &$\left<u_r^2\right>/U_c^2$ &   0.011 &    0.57 & 1.00 \\
  $F_3$ &$\left<u_\theta^2\right>/U_c^2$ &   0.009 &   0.62  & 0.84 \\
  $F_4$ & $k/U_c^2$ &   0.045 &  0.71  & 0.68 \\
  $F_5$ & $\left<\epsilon\right>b_g/U_c^3$ &   0.004 & 0.76  & 0.72 \\
 
  \end{tabular}
  \end{center}
  \caption{Coefficients of double-Gaussian function in equation \ref{Apeq:1} used to fit various parameters measured in the experiment}
  \label{Aptab:1}
  \end{table}

\noindent
The in-plane shear-stress is fitted into a polynomial-Gaussian function in equation \ref{Apeq:2}.
\begin{equation}
\centering
S = \left<u_zu_r\right>/U_c^2 = \left(0.0124\eta + 0.0286\eta^3 - 0.0113\eta^5\right)\mathrm{exp}\left(-1.47\eta^2\right)
\label{Apeq:2}
\end{equation}

\noindent
The axial and radial transport of the turbulent kinetic energy ($k$) are fitted into the polynomial-Gaussian functions in equations \ref{Apeq:3} and \ref{Apeq:4}, respectively.
\begin{equation}
\centering
T_1 = \left<u_zk\right>/U_c^3 = \left(-0.0016 - 0.0252\eta^2 -0.0181\eta^4 + 0.0746\eta^6\right)\mathrm{exp}\left(-2.62\eta^2\right)
\label{Apeq:3}
\end{equation}

\begin{equation}
\centering
T_2 = \left<u_rk\right>/U_c^3 = \left(-0.0008\eta  0.0048\eta^3 - 0.0139\eta^5 - 0.0049\eta^7\right)\mathrm{exp}\left(-1.76\eta^2\right)
\label{Apeq:4}
\end{equation}

\section{}
\label{appx2}
Expressions for the advection ($A$), shear-production ($P_s$), and turbulent tranport ($T$) terms in cylindrical co-ordinate system in the turbulent kinetic energy budget are summarized below. Each term is expressed in terms of the similarity variable $\eta=r/b_g$.

\begin{itemize}
\item \textbf{Advection term}
\noindent
\begin{equation*}
\centering
A = -[\left<U_z\right>\frac{\partial k}{\partial z} + \left<U_r\right>\frac{\partial k}{\partial r}]
\end{equation*}

\vspace{20pt}
\begin{equation}
\centering
A \times \frac{b_g}{U_c^3}= \left[\frac{2}{3} F_4(\eta) + \eta\frac{d F_4(\eta)}{d \eta} \right] \beta e^{-\eta^2} - \left[ G(\eta)\frac{d F_4(\eta)}{d\eta} \right]
\end{equation}

\noindent
\item \textbf{Production term}
\begin{equation*}
\centering
P = -\left[\underbrace{\left<{u_z^2}\right>\frac{\partial \left<U_z\right>}{\partial z}}_{P_1}+
\underbrace{\left<{u_r^2}\right>\frac{\partial \left<U_r\right>}{\partial r}}_{P_2} + \underbrace{\frac{\left<U_r\right>\left<{u_\theta^2}\right>}{r}}_{P_3} + 
\underbrace{\left<{u_zu_r}\right>\left( \frac{\partial \left<U_z\right>}{\partial r} +  \frac{\partial \left<U_r\right>}{\partial z} \right)}_{P_4}\right]
\end{equation*}

\begin{align}
P_1 \times \frac{b_g}{U_c^3} &= \frac{\beta}{3} F_1(\eta) e^{-\eta^2}(1-6\eta^2)\\
P_2 \times \frac{b_g}{U_c^3} &= - F_2(\eta)\frac{d G(\eta)}{d \eta}\\
P_3 \times \frac{b_g}{U_c^3} &=-\frac{1}{\eta}F_3(\eta)G(\eta)\\
P_4\times \frac{b_g}{U_c^3} &= S(\eta)\left( 2\eta e^{-\eta^2} + \frac{\beta}{3}G(\eta)+ \eta \beta \frac{dG(\eta)}{d\eta}\right)
\end{align}

\noindent
\item \textbf{Turbulent Transport}
\begin{equation*}
\centering
T = -\left[\frac{\partial \left<{u_zk}\right>}{\partial z} + \frac{1}{r}\frac{\partial \left<{u_rk}\right>}{\partial r}\right]
\end{equation*}

\begin{equation}
\centering
T \times \frac{b_g}{U_c^3} =  \beta\left( T_1(\eta) + \eta\frac{dT1}{d\eta}\right)  -\left( \frac{T_2(\eta)}{\eta} + \frac{\partial T_2(\eta)}{\partial \eta}\right)
\end{equation}

\end{itemize}

\bibliographystyle{jfm}
\bibliography{Reference}

\begin{thebibliography}{49}
\expandafter\ifx\csname natexlab\endcsname\relax\def\natexlab#1{#1}\fi
\def\au#1{#1} \def\ed#1{#1} \def\yr#1{#1}\def\at#1{#1}\def\jt#1{\textit{#1}}
  \def\bt#1{#1}\def\bvol#1{\textbf{#1}} \def\vol#1{#1} \def\pg#1{#1}
  \def\publ#1{#1}\def\arxiv#1{#1}\def\org#1{#1}\def\st#1{\textit{#1}}

\bibitem[Almeras {\em et~al.\/}(2017)Almeras, Mathai, Lohse \&
  Sun]{Almeras2017}
{\sc \au{Almeras, E.}, \au{Mathai, V.}, \au{Lohse, D.} \& \au{Sun, C.}}
  \yr{2017}  \at{Experimental investigation of the turbulence induced by a
  bubble swarm rising within incident turbulence}.  \jt{J. Fluid Mech.}
  \bvol{825},  \pg{1091--1112}.

\bibitem[Baines \& Sparks(2005)]{Baines2005}
{\sc \au{Baines, P.} \& \au{Sparks, R.}} \yr{2005}  \at{Dynamics of giant
  volcanic ash clouds from supervolcanic eruptions}.  \jt{Geophys. Res. Lett.}
  \bvol{32},  \pg{L24808}.

\bibitem[Baines(2008)]{Baines2008}
{\sc \au{Baines, P.~G.}} \yr{2008}  \at{Mixing in downslope flows in the ocean
  ‐ plumes versus gravity currents}.  \jt{Atmosphere-Ocean}  \bvol{46},
  \pg{402--419}.

\bibitem[Balachandar \& Eaton(2010)]{Balachandar2010}
{\sc \au{Balachandar, S.} \& \au{Eaton, J.~K.}} \yr{2010}  \at{Turbulent
  dispersed multiphase flow}.  \jt{Annual Review of Fluid Mechanics}
  \bvol{42},  \pg{111--133}.

\bibitem[Bombardelli {\em et~al.\/}(2007)Bombardelli, Buscaglia, Rehmann,
  Rincon \& Garcia]{Bombardelli2007}
{\sc \au{Bombardelli, F.A.}, \au{Buscaglia, G.C.}, \au{Rehmann, C.R.},
  \au{Rincon, L.E.} \& \au{Garcia, M.H.}} \yr{2007}  \at{Modeling and scaling
  of aeration bubble plumes: A two-phase ow analysis}.  \jt{J. Hydraul. Res.}
  \bvol{45}~(5),  \pg{617--630}.

\bibitem[Bordoloi \& Variano(2017)]{Bordoloi2017}
{\sc \au{Bordoloi, A.~D.} \& \au{Variano, E.~A.}} \yr{2017}  \at{Rotational
  kinematics of large cylindrical particles in turbulence}.  \jt{J. Fluid
  Mech.}  \bvol{815},  \pg{199--222}.

\bibitem[Cartellier \& Rivi\`{e}re(2001)]{Cartellier2001}
{\sc \au{Cartellier, A.} \& \au{Rivi\`{e}re, N.}} \yr{2001}  \at{Bubble-induced
  agitation and microstructure in uniform bubbly flows at small to moderate
  particle reynolds numbers}.  \jt{Physics of Fluids}  \bvol{13},  \pg{2165}.

\bibitem[Charonko \& Prestridge(2017)]{Charonko2017}
{\sc \au{Charonko, J.} \& \au{Prestridge, K.}} \yr{2017}  \at{Variable-density
  mixing in turbulent jets with co-flow}.  \jt{J. Fluid Mech.}  \bvol{825},
  \pg{887--921}.

\bibitem[Clift {\em et~al.\/}(1978)Clift, Grace \& Weber]{Clfit1978}
{\sc \au{Clift, R.}, \au{Grace, J.~R.} \& \au{Weber, M.E.}} \yr{1978} {\em
  Bubbles, drops and particles\/}.  \publ{New York: Academic Press, Inc.}

\bibitem[Darisse {\em et~al.\/}(2012)Darisse, Lemay \&
  Bena\''{i}ssa]{Darisse2012}
{\sc \au{Darisse, A.}, \au{Lemay, J.} \& \au{Bena\''{i}ssa, A.}} \yr{2012}
  \at{Ldv measurements of well converged third order moments in the far field
  of a free turbulent round jet}.  \jt{Experimental Thermal and Fluid Science}
  \bvol{44}~(213),  \pg{823--833}.

\bibitem[Darisse {\em et~al.\/}(2015)Darisse, Lemay \& Benaïssa]{Darisse2015}
{\sc \au{Darisse, Alexis}, \au{Lemay, Jean} \& \au{Benaïssa, Azemi}} \yr{2015}
   \at{Budgets of turbulent kinetic energy, reynolds stresses, variance of
  temperature fluctuations and turbulent heat fluxes in a round jet}.
  \jt{Journal of Fluid Mechanics}  \bvol{774},  \pg{95--142}.

\bibitem[Duncan {\em et~al.\/}(2009)Duncan, Seol \& Socolofsky]{Duncan2009}
{\sc \au{Duncan, B.~B.}, \au{Seol, D-G.} \& \au{Socolofsky, S.~A.}} \yr{2009}
  \at{Quantication of turbulence properties in bubble plumes using vortex
  identication methods}.  \jt{Phys. Fluids}  \bvol{21}~(7),  \pg{0}.

\bibitem[Fraga \& Stoesser(2016)]{Fraga2016}
{\sc \au{Fraga, Bru{\~{n}}o} \& \au{Stoesser, Thorsten}} \yr{2016}
  \at{Influence of bubble size, diffuser width, and flow rate on the integral
  behavior of bubble plumes}.  \jt{Journal of Geophysical Research: Oceans}
  \bvol{121}~(6),  \pg{3887--3904}.

\bibitem[Freeth(1987)]{Freeth1987}
{\sc \au{Freeth, S.}} \yr{1987}  \at{The lake nyos disaster}.  \jt{Nature}
  \bvol{325},  \pg{104--5}.

\bibitem[Garc\'{i}a \& Garc\'{i}a(2006)]{Garcia2006}
{\sc \au{Garc\'{i}a, C.~M.} \& \au{Garc\'{i}a, M.~H.}} \yr{2006}
  \at{Characterization of flow turbulence in large-scale bubble-plume
  experiments}.  \jt{Experiments in Fluids}  \bvol{41}~(1),  \pg{91--101}.

\bibitem[George \& Hussein(1991)]{George1991}
{\sc \au{George, W.~K.} \& \au{Hussein, H.}} \yr{1991}  \at{Locally
  axisymmetric turbulence}.  \jt{J. Fluid Mech.}  \bvol{233},  \pg{1}.

\bibitem[Gonzalez \& Woods(2007)]{Gonzalez2007}
{\sc \au{Gonzalez, Rafael~C.} \& \au{Woods, Richard~E.}} \yr{2007} {\em Digital
  Image Processing (3rd Edition)\/}.  \publ{Pearson}.

\bibitem[Guazzelli \& Morris(2012)]{Guazzelli2012}
{\sc \au{Guazzelli, E.} \& \au{Morris, J.~F.}} \yr{2012} {\em A physical
  introduction to suspension dynamics\/}.  \publ{Cambridge University Press,
  Cambridge; New York}.

\bibitem[Huppert \& Neufeld(2014)]{Huppert2014}
{\sc \au{Huppert, H.~E.} \& \au{Neufeld, J.~A.}} \yr{2014}  \at{The fluid
  mechanics of carbon dioxide sequestration}.  \jt{Annual Review of Fluid
  Mechanics}  \bvol{46},  \pg{255--272}.

\bibitem[Kataoka \& Serizawa(1989)]{Kataoka1989}
{\sc \au{Kataoka, I.} \& \au{Serizawa, A.}} \yr{1989}  \at{Basic equations of
  turbulence in gas–liquid two-phase flow}.  \jt{Intl J. Multiphase Flow}
  \bvol{15},  \pg{843–855}.

\bibitem[Lai {\em et~al.\/}(2018)Lai, Fraga, Chan \& Dodd]{Lai2018a}
{\sc \au{Lai, C. C.~K.}, \au{Fraga, B.}, \au{Chan, W. R.~H.} \& \au{Dodd,
  M.~S.}} \yr{2018}  \bt{Energy cascade in a homogeneous swarm of bubbles
  rising in a vertical channel}. techreport.  \org{Center of Turbulence
  Research, Proceedings of Summer Program}.

\bibitem[Lai \& Socolofsky(2018)]{Lai2018}
{\sc \au{Lai, C. C.~K.} \& \au{Socolofsky, S.~A.}} \yr{2018}  \at{Budgets of
  turbulent kinetic energy, reynolds stresses, and dissipation in a turbulent
  round jet discharged into a stagnant ambient}.  \jt{Environmental Fluid
  Mechanics}  \pg{p. 1–29}.

\bibitem[Lai \& Socolofsky(2019)]{Lai2019}
{\sc \au{Lai, C. C.~K.} \& \au{Socolofsky, S.~A.}} \yr{2019}  \at{The turbulent
  kinetic energy budget in a bubble plume}.  \jt{Journal of Fluid Mechanics}
  \bvol{865},  \pg{993--1041}.

\bibitem[Lance(1991)]{Lance1991}
{\sc \au{Lance, M. \&~Bataille, J.}} \yr{1991}  \at{Turbulence in the liquid
  phase of a uniform bubbly air-water flow}.  \jt{J. Fluid Mech.}  \bvol{222},
  \pg{95--118}.

\bibitem[Lee \& Chu(2003)]{Lee2003}
{\sc \au{Lee, J. H.~W.} \& \au{Chu, V.~H.}} \yr{2003} {\em Turbulent Jets and
  Plumes – a Lagrangian Approach\/}.  \publ{Kluwer Academic}.

\bibitem[Lumley(1978)]{Lumley1978}
{\sc \au{Lumley, J.~L.}} \yr{1978}  \at{Computational modeling of turbulent
  flows}.  \jt{Adv. Appl. Mech.}  \bvol{18},  \pg{123--176}.

\bibitem[Mercado {\em et~al.\/}(2010)Mercado, Gomes, van Gils \&
  Lohse]{Mercado2010}
{\sc \au{Mercado, J.M.}, \au{Gomes, D.C.}, \au{van Gils, D.a dn~Sun, C.} \&
  \au{Lohse, D.}} \yr{2010}  \at{On bubble clustering and energy spectra in
  pseudo-turbulence}.  \jt{J. Fluid Mech.}  \bvol{650},  \pg{287--306}.

\bibitem[Milgram(1983)]{Milgram1983}
{\sc \au{Milgram, J.~H.}} \yr{1983}  \at{Mean flow in round bubble plumes}.
  \jt{Journal of Fluid Mechanics}  \bvol{133},  \pg{345–376}.

\bibitem[Morton {\em et~al.\/}(1956)Morton, Taylor \& Turner]{Morton1956}
{\sc \au{Morton, B.}, \au{Taylor, G.} \& \au{Turner, J.}} \yr{1956}
  \at{Turbulent gravitational convection from maintained and instantaneous
  sources}.  \jt{Proceedings of the Royal Society of London Series
  a-Mathematical and Physical Sciences}  \bvol{234}~(1),  \pg{1196}.

\bibitem[Papanicolaou \& List(1988)]{Papanicolaou1988}
{\sc \au{Papanicolaou, P.N.} \& \au{List, E.J.}} \yr{1988}  \at{Investigations
  of round vertical buoyant jets}.  \jt{J. Fluid Mech.}  \bvol{195},
  \pg{341--395}.

\bibitem[Pope(2000)]{Pope2000}
{\sc \au{Pope, S.~B.}} \yr{2000} {\em Turbulent flows\/}.  \publ{Cambridge
  University Press, Cambridge}.

\bibitem[Prakash {\em et~al.\/}(2016)Prakash, Mercado, Wijngaarden, Mancilla,
  Tagawa, Lohse \& Sun]{Prakash2016}
{\sc \au{Prakash, V.~N.}, \au{Mercado, J.~M.}, \au{Wijngaarden, E.~M.},
  \au{Mancilla, E.}, \au{Tagawa, Y.}, \au{Lohse, D.} \& \au{Sun, C.}} \yr{2016}
   \at{Energy spectra in turbulent bubbly flows}.  \jt{J. Fluid Mech.}
  \bvol{791},  \pg{174--190}.

\bibitem[Riboux {\em et~al.\/}(2013)Riboux, Legendre \& Risso]{Riboux2013}
{\sc \au{Riboux, G.}, \au{Legendre, D.} \& \au{Risso, F.}} \yr{2013}  \at{A
  model of bubble-induced turbulence based on large-scale wake interactions.}
  \jt{J. Fluid Mech.}  \bvol{719},  \pg{362--387}.

\bibitem[Riboux {\em et~al.\/}(2010)Riboux, Risso \& Legendre]{Riboux2010}
{\sc \au{Riboux, G.}, \au{Risso, F.} \& \au{Legendre, D.}} \yr{2010}
  \at{Experimental characterization of the agitation generated by bubbles
  rising at high reynolds number.}  \jt{J. Fluid Mech.}  \bvol{643},
  \pg{509--539}.

\bibitem[Risso(2018)]{Risso2018}
{\sc \au{Risso, F.}} \yr{2018}  \at{Agitation, mixing, and transfers induced by
  bubbles}.  \jt{Annu. Rev. Fluid Mech.}  \bvol{50},  \pg{25--48}.

\bibitem[Risso \& Ellingsen(2002)]{Risso2002}
{\sc \au{Risso, F.} \& \au{Ellingsen, K.}} \yr{2002}  \at{Velocity uctuations
  in a homogeneous dilute dispersion of high-reynolds-number rising bubbles}.
  \jt{J. Fluid Mech.}  \bvol{453},  \pg{395--410}.

\bibitem[Santarelli {\em et~al.\/}(2016)Santarelli, Rouseel \&
  Frohlich]{Santarelli2016}
{\sc \au{Santarelli, C.}, \au{Rouseel, J.} \& \au{Frohlich}} \yr{2016}
  \at{Budget analysis of the turbulent kinetic energy for bubbly flow in a
  vertical channel}.  \jt{Chemical Engineering Science}  \bvol{141},
  \pg{46--62}.

\bibitem[Seol {\em et~al.\/}(2009)Seol, Duncan \& Socolofsky]{Seol2009}
{\sc \au{Seol, D-G.}, \au{Duncan, B.B.} \& \au{Socolofsky, S.A.}} \yr{2009}
  \at{Measurements of behavioral properties of entrained ambient water in
  stratied bubble plume}.  \jt{J. Hyd. Eng., ASCE}  \bvol{135}~(11),
  \pg{983--988}.

\bibitem[Simiano {\em et~al.\/}(2009)Simiano, Lakehal, Lance \&
  Yadigaroglu]{Simiano2009}
{\sc \au{Simiano, M.}, \au{Lakehal, D.}, \au{Lance, M.} \& \au{Yadigaroglu}}
  \yr{2009}  \at{Turbulent transport mechanisms in oscillating bubble plumes}.
  \jt{J. Fluid Mech.}  \bvol{633},  \pg{191--231}.

\bibitem[Socolofsky {\em et~al.\/}(2011)Socolofsky, Adams \&
  Sherwood]{Socolofsky2011}
{\sc \au{Socolofsky, S.~A.}, \au{Adams, E.~E.} \& \au{Sherwood, C.~R.}}
  \yr{2011}  \at{Formation dynamics of subsurface hydrocarbon intrusions
  following the deepwater horizon blowout}.  \jt{Geophysical Research Letters}
  \bvol{38}~(9),  \pg{L09602}.

\bibitem[Socolofsky \& Bhaumik(2008)]{Socolofsky2008}
{\sc \au{Socolofsky, S.~A.} \& \au{Bhaumik, T.}} \yr{2008}  \at{Dissolution of
  direct ocean carbon sequestration plumes using an integral model approach}.
  \jt{Journal of Hydraulic Engineering-ASCE}  \bvol{134}~(11),
  \pg{1570–1578}.

\bibitem[Soga \& Rehmann(2004)]{Soga2004}
{\sc \au{Soga, C.} \& \au{Rehmann, C.}} \yr{2004}  \at{Dissipation of turbulent
  kinetic energy near a bubble plume}.  \jt{Journal of Hydraulic Engineering}
  \bvol{130}~(5),  \pg{441– 449}.

\bibitem[Sun \& Faeth(1986)]{Sun1986}
{\sc \au{Sun, T.-Y.} \& \au{Faeth, G.~M.}} \yr{1986}  \at{Structure of
  turbulent bubbly jets—ii. phase property profiles}.  \jt{International
  Journal of Multiphase Flow}  \bvol{12}~(1),  \pg{115--126}.

\bibitem[Tan \& Huang(2015)]{Tan2015}
{\sc \au{Tan, C.-Y.} \& \au{Huang, Y.-X.}} \yr{2015}  \at{Dependence of
  refractive index on concentration and temperature in electrolyte solution,
  polar solution, nonpolar solution, and protein solution}.  \jt{J. Chem. Eng.
  Data}  \bvol{60}~(10),  \pg{2827–2833}.

\bibitem[Wain \& Rehmann(2005)]{Wain2005}
{\sc \au{Wain, D.~J.} \& \au{Rehmann, C.~R.}} \yr{2005}  \at{Eddy diffusivity
  near bubble plumes}.  \jt{Water Resources Research}  \bvol{41}~(9).

\bibitem[Wang {\em et~al.\/}(2016)Wang, Socolofsky, Breier \&
  Seewald]{Wang2016}
{\sc \au{Wang, B.}, \au{Socolofsky, S.A.}, \au{Breier, J.A.} \& \au{Seewald,
  J.S.}} \yr{2016}  \at{Observations of bubbles in natural seep ares at mc 118
  and gc 600 using in situ quantitative imaging}.  \jt{J. Geophys. Res. Oceans}
   \bvol{121},  \pg{2203--2230}.

\bibitem[Wang \& Law(2002)]{Wang2002}
{\sc \au{Wang, H.} \& \au{Law, A.W.K.}} \yr{2002}  \at{Second-order integral
  model for a round turbulent buoyant jet}.  \jt{J. Fluid Mech.}  \bvol{459},
  \pg{3}.

\bibitem[Westerweel \& Scarano(2005)]{Westerweel2005}
{\sc \au{Westerweel, Jerry} \& \au{Scarano, Fulvio}} \yr{2005}  \at{Universal
  outlier detection for {PIV} data}.  \jt{Experiments in Fluids}
  \bvol{39}~(6),  \pg{1096--1100}.

\bibitem[Woods(2010)]{Woods2010}
{\sc \au{Woods, A.~W.}} \yr{2010}  \at{Turbulent plumes in nature}.  \jt{Annu.
  Rev. Fluid Mech.}  \bvol{42},  \pg{391–412}.

\end{thebibliography}

\end{document}